\documentclass{aastex}
\usepackage{graphicx,times}             
\usepackage{longtable}

\shorttitle{Cosmic Evolution of Long Gamma-Ray Burst Luminosity}
\shortauthors{Deng et al.} \slugcomment{The Astrophysical Journal, 000:000-000 2015 00 00}
\begin{document}
\title{Cosmic Evolution of Long Gamma-Ray Burst Luminosity}
\author{Can-Min Deng \altaffilmark{1,2}, Xiang-Gao Wang  \altaffilmark{1,2}, Bei-Bei Guo\altaffilmark{1,2}, Rui-Jing Lu \altaffilmark{1,2}, Yuan-Zhu Wang \altaffilmark{1,2}, Jun-Jie Wei \altaffilmark{3,4}, Xue-Feng Wu\altaffilmark{3,4}, and En-Wei Liang \altaffilmark{1,2}}
  \altaffiltext{1}{GXU-NAOC Center for Astrophysics and Space Sciences, Department of Physics, Guangxi University, Nanning 530004, China;luruijing@gxu.edu.cn; lew@gxu.edu.cn}
  \altaffiltext{2}{Guangxi Key Laboratory for the Relativistic Astrophysics, Nanning 530004, China}
  \altaffiltext{3}{Purple Mountain Observatory, Chinese Academy of Sciences, Nanjing 210008, China}
  \altaffiltext{4}{Graduate University of Chinese Academy of Sciences, Beijing 100049, China}
\begin{abstract}
The cosmic evolution of gamma-ray burst (GRB) luminosity is essential for revealing the GRB physics and for using GRBs as cosmological probes. We investigate the luminosity evolution of long GRBs with a large sample of 258 {\em Swift}/BAT GRBs. Parameterized the peak luminosity of individual GRBs evolves as $L_{\rm p}\propto{\rm }(1+z)^{k}$, we get $k=1.49\pm0.19$ using the non-parametric $\tau$ statistics method without considering observational biases of GRB trigger and redshift measurement. By modeling these biases with the observed peak flux and characterizing the peak luminosity function of long GRBs as a smoothly broken power-law with a break that evolves as $L_{\rm b}\propto (1+z)^{k_{\rm b}}$, we obtain $k_{\rm b}=1.14^{+0.99}_{-0.47}$  through simulations based on assumption that the long GRB rate follows the star formation rate (SFR) incorporating with cosmic metallicity history. The derived $k$ and $k_b$ values are systematically smaller than that reported in previous papers. By removing the observational biases of the GRB trigger and redshift measurement based on our simulation analysis, we generate mock {\em complete} samples of 258 and 1000 GRBs to examine how these biases affects on the $\tau$ statistics method. We get $k=0.94\pm 0.14$ and $k=0.80\pm 0.09$ for the two samples, indicating that these observational biases may lead to overestimate the $k$ value. With the large uncertain of $k_b$ derived from our simulation analysis, one even cannot convincingly argue a robust evolution feature of the GRB luminosity.
\end{abstract}
\section{Introduction}           
\label{sect:intro}
Gamma-ray bursts (GRBs) are the most luminous events in the universe. They have been detected up to a redshift of 9.4 so far (GRB 090429B; Cucchiara et al. 2011). It is generally believed that these events are from mergers of compact stellar binaries (Type I) or core collapses of massive stars (Type II) (e.g. Woosley 1993; Paczynski 1998; Woosley \& Bloom 2006;  Zhang 2006, Zhang et al. et al. 2007, 2009). The durations of prompt gamma-rays of Type I GRBs are usually short and they are long for Type II GRBs. This may lead to a tentative bimodal distribution of the burst duration and two groups of GRBs were proposed, i.e., short GRBs (SGRBs) with $T_{90}<2$ s and long GRBs (LGRBs) with $T_{90}>2$ s, where $T_{90}$ is the time interval of from 5\% to 95\% gamma-ray photons collected by a given instrument (Kouveliotou et al. 1993)\footnote{Note that the statistical significance of the bimodal distribution depends on instruments and energy bands (e.q., Qin et al. 2013). Burst duration only is difficult to clarify the physical origin of some GRBs, such as GRBs 050724 and 060614 and new classification scheme and classification methods have been proposed ( Zhang 2006; Zhang et al. 2007, 2009; L\"{u} et al. 2010).}. Since the LGRB rate may follow the cosmic star formation rate (SFR), it has been proposed that LGRBs could be probes for the high-redshift universe, and may be used as a potential tracer of SFR at high redshift as where it becomes difficult for other methods (Totani 1997; Wijers et al. 1998; Blain \& Natarajan 2000; Lamb \& Reichart 2000; Porciani \& Madau 2001; Piran 2004; Zhang \& M\'{e}sz\'{a}ros 2004; Zhang 2007; Robertson \& Ellis 2012; Elliott et al. 2012; Wei et al. 2014; wang et al. 2015). It was also proposed that LGRBs may be promising rulers to measure cosmological parameters and dark energy (e.g., Dai et al. 2004, Ghirlanda et al. 2004; Liang \& Zhang et al. 2005).

The cosmic evolution of GRB luminosity is essential for revealing the GRB physics and using GRBs as cosmological probes, but it is poorly known being due to lack of a large and complete GRB sample in redshift. With growing sample of high-redshift LGRBs since the launch of {\em Swift} satellite, it was found that GRBs rate increases significantly faster than the SFR, especially at high-$z$ (Daigne et al. 2006; Le \& Dermer 2007; Salvaterra \& Chincarini 2007; Guetta \& Piran 2007; Y\"{u}ksel et al. 2008; Li 2008; Salvaterra et al. 2008; Kistler et al. 2008).
It is unclear whether this excess is due to some sort of evolutions in an intrinsic luminosity/mass function (Salvaterra \& Chincarini 2007; Salvaterra et al. 2009, 2012) or cosmic evolution of GRB rate (e.g., Butler et al. 2010; Qin et al. 2010; Wanderman \& Prian 2010; Wang \& Dai 2011). Coward et al. (2013) even proposed that it is not necessary to invoke luminosity evolution with redshift to explain the observed GRB rate at high-z by carefully taking selection effects into account (see also Howell \& Coward, 2013)

Attempts to measure the cosmic evolution of LGRB luminosity have been made by several groups. The non-parametric $\tau$ statistical method (Efron \& Petrosian 1992) is usually used to estimate the possible intrinsic $L-z$ correlation by simplifying the evolution feature as $L_{\rm p}=L_{\rm p, 0}(1+z)^{k}$ for individual GRBs, where the $L_{p,0}$ is the luminosity of local GRBs. Strong $L-z$ dependence with a $k$ value varying from 1.4 to 2.7 with different samples is reported (Lloyd-Ronning et al. 2002; Yonetoku et al. 2004; Kocevski \& Liang 2006; Petrosian et al. 2009). Alternately, the comic evolution of GRB luminosity was also estimated with the cosmic evolution of their luminosity function. With this approach, the luminosity function is usually adopted as a broken power-law with a break at $L_{\rm b}$, which evolves as $L_{\rm b}=L_{b, 0}(1+z)^{k_{\rm b}}$. $k_b$ then can be estimated by fitting the observed $L_{\rm p}$ and $z$ distributions via Monte Carlo simulations under assumption that GRB rate follows the SFR. Using this approach, the $k_b$ derived from different samples is around 2 (Salvaterra et al. 2012; Tan et al. 2013).

Sample selection and observational biases are critical for measuring the GRB luminosity evolution feature. As mentioned above, results derived from different samples are significantly different. The redshift-known GRB samples are inevitably suffered observational biases on the flux truncation, trigger probability, redshift measurement, etc (e.g., Coward et al. 2013). By modeling these biases with a large and uniform sample of GRBs with redshift measurement in a broad redshift range is essential to robustly estimate the luminosity evolution. The {\em Swift} mission has established a considerable sample of LGRBs with redshift measurement in a range of $0.1\sim 9.4$ over 10 operation years. This paper revisits the cosmic luminosity evolution with the current redshift-known GRB sample by carefully considering the possible observational biases. We report our sample and the apparent luminosity dependence to redshift in \S 2. We derive the $L-z$ dependence for GRBs from the current sample of Swift GRBs with redshift measurement with the non-parametric $\tau$-statistics method (\S 3). By considering the BAT trigger probability and redshift measurement probability, we evaluate the $L-z$ dependence with simulations by assuming that GRB rate follows the SFR incorporating with the cosmic metallicity history (\S 4). Based on our simulation analysis, we further investigate how these observational biases affect the results of the $\tau$ statistics method (\S 5). We present our conclusions and make brief discussion in \S 6. Throughout the paper the cosmological parameters $H_{0}=71$ km s$^{-1}$ Mpc$^{-1}$, $\Omega_{M}=0.3$ and $\Omega_{\Lambda}=0.7$ are adopted.

\section{Sample selection and Data}
\label{sect:Obs}
Our sample includes only the redshift-known long GRBs observed with Swift/BAT from Jan. 2005 to April 2015. Low-luminosity GRBs are excluded since they may belong to another distinct population (e.g., Liang et al. 2007; Chapman et al. 2007; L\"{u} et al. 2010). Although the durations of some GRBs, such as GRBs 050724 and 060614, are larger than 2 seconds, we do not include them in our sample since they may be from merger of compact stars (Berger et al. 2005; Gehrels et al. 2006; Zhang et al. 2007; 2009). The afterglow and host galaxy observations of the high-z short GRB 090426 ($T_{90}=1.2$s) show that it may be from collapse of a massive star (Antonelli et al. 2009; Xin et al. 2010). We thus include this GRB in our sample. We finally obtain a sample of 258 GRBs. They are reported in Table 1. Their redshifts ($z$), peak photon fluxes ($P$), and photon indices ($\Gamma$) are taken from the NASA website \footnote{http://swift.gsfc.nasa.gov/}.

GRB radiation spectra are very broad. They can be well fitted with the so-called Band function,
which is a smooth broken power-law characterized with low and high photon indices at a break energy (Band et al. 1993).
The peak energy ($E_{\rm p}$) of the $\nu f_\nu$ spectrum ranges from several keVs to MeVs (Preece et al. 2000;
Liang \& Dai 2004; L\"{U} et al. 2010; Zhang et al. 2011; von Kienlin et al. 2014).
BAT energy band covers a very narrow range, i.e., 15-150 keV.
The GRB spectra observed with BAT are adequately fitted with a single power-law, $N(E)\propto E^{-\Gamma}$ (Zhang et al. 2007; Sakamoto et al. 2009).
Since lack of $E_{\rm p}$ information, we do not calculate the bolometric luminosity of the GRBs in order to avoid unreasonable extrapolation
with spectral information derived from BAT data. We calculate their peak luminosity ($L_{\rm p}$) using the 1s peak fluxes measured
 in the 15-150 keV band for our analysis. The distribution of the GRBs in the $\log L_{\rm p}-\log (1+z)$ plane is shown in Figure 1. A tentative
 correlation with large scatter is observed. The Spearman correlation analysis yields a correlation efficient of 0.72
  and a chance probability $p<10^{-4}$. With the ordinary least squares bisector algorithm and consider an intrinsic scatter
  of $\sigma_{\rm in}$ (e.g., Weiner et al. 2006), we derive a relation of  $\log L_{\rm p}=(49.98\pm 0.09)+(2.95\pm 0.19)\log[(1+z)]$ and $\sigma_{\rm in}=0.72\pm0.02$ by taking the luminosity uncertainty into account\footnote{Since the errors are very small and even no uncertainty is reported for the redshifts of most GRBs, we do not take into account the uncertainties of redshifts in our fit.}.

The flux threshold of Swift/BAT is complicated, and the trigger probability of a GRB with a peak flux close to the instrument threshold is much lower than that of high-flux GRBs (e.g., Stern et al. 2001 Qin et al. 2010). In addition, in the image trigger mode, a GRB trigger also depends on the burst duration, hence the burst fluence (Band 2006, Sakamoto et al. 2007, Virgili et al. 2009). We adopt a flux threshold as $1.0 \times 10^{-8}$ erg cm$^{-2}$ s$^{-1}$, which is available by the BAT team in the NASA web page\footnote{ {\em http://swift.gsfc.nasa.gov/about\_swift/bat\_desc.html}}.It is also shown in Figure 1.

\section{Measuring the intrinsic $L_{\rm p}-z$ dependence with the $\tau$ statistics method}
Because the sample is greatly suffered from flux truncation effect, it is uncertain whether the apparent $L_{\rm p}-z$ relation is intrinsic or only due to the truncation effect. The non-parametric $\tau$ statistical method is a simple, straightforward way to estimate the truncation effect on an observed correlation (Efron \& Petrosian 1992; see also Lloyd-Ronning et al. 2002, Yonetoku et al. 2004 and Kocevski \& Liang 2006). We first apply this method to measure the intrinsic dependence of $L_{\rm p}$ to $(1+z)$ for our sample. We outline this technique as following.

Firstly, we pick up an associated set of ($L_{i}$, $z_{i}$),
\begin{equation}\label{associatedset}
J_{i}=\{j~|L_{j}>L_{i},~ z_{j}<z_{\rm i,lim}\}, ~~~~~~  {\rm for }~~ 1\leq i \leq 258 \;\;,
\end{equation}
where $z_{\rm i,lim}$ is the redshift corresponding to the flux threshold for a GRB of ($L_{i}, z_{i}$), and $N$ is the size of our sample. The number of the set \{$J_{i}$\} is defined as $N_{i}$. Independence between $L_{i}$ and $z_{i}$ would make the number of the sample
\begin{equation}\label{Rset}
R_{i}={\rm number} \{j \in J_{i} ~| z_{j}\leq z_{i}\},
\end{equation}
uniformly distribute between 1 and $N_{i}$. To quantify the correlation degree between the two quantities, the test statistic $\tau$ is introduced as
\begin{equation}\label{tau}
\tau=\frac{\sum_{i}(R_{i}-E_{i})}{\sqrt{\sum_{i} V_{i}}},
\end{equation}
where $E_{i}=(N_{i}+1)/2$ and $V_{i}=(N_{i}^2-1)/12$ are the expected mean and variance for the uniform distribution, respectively. This $\tau$ value is employed to measure the data correlation degree between the two quantities of $L$ and $z$ in units of standard deviation. In case of that the two quantities are intrinsically independent without any correlation, the $\tau$ statistics gives $\tau=0$.  Note that $N_{i}$ should be greater than 5 since this method assumes Gaussian statistics. For the details of this technique please refer to Efron \& Petrosian (1992).

We simply parameterize the cosmic evolution of the GRB luminosity as $L =L_0(1+z)^{k}$ for individual GRBs (see also Lloyd-Ronning et al. 2002; Yonetoku et al. 2004). We vary the $k$ value and derive the corresponding $\tau$ value. Figure 2 shows $\tau$ as a function of $k$ for our GRB sample. It is found that $|\tau|=7.7$ at $k=0$, indicating that the null hypothesis that $L$ and $z$ are statistically independent is rejected in a confidence level of $7.7 \sigma$. In the case of $k=1.49\pm0.19$, we have $\tau=0$ within 1 $\sigma$ significance. Therefore, the intrinsic dependence of $L_{\rm p}$ to $1+z$ is $L_{\rm p}\propto (1+z)^{1.49\pm0.19}$. It is much shallower than the apparent one, which is $L_{\rm p}\propto (1+z)^{2.95\pm0.19}$. The apparent $\log L_{\rm p}-\log(1+z)$ relation is dominated by the instrument selection effect.

\section{Measuring the intrinsic $L_{\rm p}-z$ dependence with the comics evolution of the luminosity function by Monte Carlo simulations}
\label{sect:MC}
The evolution of GRB luminosity can also shows up as the evolution of the GRB luminosity function. The GRB luminosity function is usually described with a smooth broken power-law (e.g., Liang et al. 2007 and references therein),
\begin{equation}\label{BPL}
\Phi(L_{\rm p})=\Phi_{p, 0}\left[\left(\frac{L}{L_{\rm b}}\right)^{\alpha_{1}
}+\left(\frac{L}{L_{\rm b}}\right)^{\alpha_{2}}\right]^{-1},
\end{equation}
where $\Phi_{\rm p, 0}$ is a normalization constant, $\alpha_{1}$ and $\alpha_{2}$ are the low and high indices of the luminosity function.
 We suggest that the evolution of GRB luminosity is parameterized with the evolution of $L_{\rm b}$ in the form of \footnote{Physically, the luminosity evolution parameterized as $L=L_0(1+z)^{k}$ for individual GRBs is comparable to that parameterized as $L_{\rm b}=L_{\rm b, 0}(1+z)^{k_b}$ for the break of the GRB luminosity function, but the derived $k_b$ and $k$ values from a given sample would not be completely the same.}
\begin{equation}
L_{\rm b}=L_{b, 0}(1+z)^{k_{\rm b}},
\end{equation}
where $L_{b,0}$ is the $L_b$ at $z=0$.

We first show how $L_b$ may evolve with redshift from the observational data. We plot the cumulative luminosity distributions in different redshift ranges in Figure 3. One can observe that they show as broken power-laws, with a larger break luminosity ($L_{\rm b}$) at higher redshift range. We fit them with Eq. \ref{BPL}. The results are reported in Table 2. One can find that $L_{\rm b}$ shows a trend that GRBs at higher redshift tends to be more luminous, but one still cannot statistically claim a correlation between the break luminosity and the redshift based on the current sample since the uncertainty of the $L_{\rm b}$ is very large, especially in the redshift intervals of above $z>1$. In the redshift interval of $0<z<1$, $L_{\rm b}$ is smaller than that in the high redshift intervals with about one order of magnitude. However, we should note that the GRB subset in this redshift interval may be contaminated by the low luminosity GRB population as we mention above.

We measure the $k_{\rm b}$ value by fitting the observed $L_{\rm p}$ and $z$ distributions in a physical context that the GRB rate follows the SFR rate incorporating the cosmic metallicity history by modeling the observational biases as reported  in Qin et al. (2010) via Monte Carlo simulations.

We assume that the GRB rate ($R_{\rm LGRB}$) traces the SFR and metallicity
history (e.g., Kistler et al. 2008; Li 2008; Qin et al. 2010; Virgili et al. 2011; Lu et al. 2012). The intrinsic GRB redshift distribution then can be written as

\begin{equation}\label{density}
\rho(z)\propto\frac{SFR(z)\Theta (\epsilon, z)}{1+z}\frac{dV(z)}{dz},
\end{equation}
 where $R_{\rm SFR}(z)$ is the star formation rate (taken the form parameterized by Y{\"u}ksel et al 2008), $\Theta (\epsilon, z)$ is the fractional mass density
belonging to metallicity below $\epsilon Z_{\bigodot}$ at a given redshift z ($Z_{\bigodot}$ is the solar metal abundance) and $\epsilon$ is determined by the metallicity threshold for the production of LGRBs (e.g. Langer \& Norman 2006). $\frac{dV(z)}{dz}$ is the co-moving volume element at redshift $z$ in a flat $\Lambda$ CDM universe, given by
\begin{equation}\label{dvdz}
\frac{dV(z)}{dz}=\frac{c}{H_0}\frac{4\pi D^2_L(z)}{(1+z)^2
[\Omega_M(1+z)^3+\Omega_\Lambda]^{1/2}},
\end{equation}
where $D_L(z)$ is the luminosity distance at $z$ and $c$ is the speed of light.

  We make Monte Carlo simulations based on Eqs. (4)--(7). The parameters of our model include $\alpha_1, \alpha_2, L_{b, 0}, k_{\rm b}$ and $\epsilon$. We set these parameters in broad ranges, i.e., $\alpha_1\in(0,2)$, $\alpha_2\in(1.8,4.5)$, $L_{b, 0}\in (10^{50},10^{53})$ erg/s, $k_{\rm b}\in(-0.5,3.0)$ and $\epsilon\in(0,1.6)$. The details of our simulation method please refer to Qin et al. (2010). We outline the simulation procedure as following.
\begin{itemize}
\item  We randomly pick up an $\epsilon$ value and generate a redshift $z$ from the probability distribution of Eq. (6), then randomly pick up a set of luminosity function parameters $\{$ $L_{\rm b, 0}$, $k_{\rm b}$, $\alpha_{1}$, $\alpha_{2} \}$ to calculate the probability distributions of $L_{\rm p}$ at $z$ with Eqs. (4) and (5). We then randomly pick up a $L_{\rm p}$ at redhsift $z$. A simulated GRB is characterized with a set of ($L_{p}$, $z$).

\item We calculate the peak flux ($F_{p}$) in 15-150 keV band for a simulated GRB, assuming a GRB photon index as $\Gamma=1.69$, which is the average value of the GRB in our BAT sample. The peak flux is compared with the BAT flux threshold to evaluate whether it can be observable with BAT.

\item We simulate the trigger probability and redshift measurement probability by modeling these probabilities as a function of the peak flux. The details please see Eqs. (9) and (10) of Qin et al. (2010).

\item We generate a mock BAT GRB sample of 258 GRBs (the same size as the observed sample) and measure the consistency between the simulated and observed GRB samples in the $\log L_p-\log (1+z)$ plane with the probability ($P_{KS}$) of the two-dimensional Kolmogorov-Smirnov test (K-S test; Press et al. 1992)\footnote{Two dimensional K-S test is not straightforward from the one-dimensional K-S test. We adopt an algorithm available in (Press et al. 1992) and  find no dependence on the ordering of the dimensions. Caveats/comments on this algorithm please see https://asaip.psu.edu/Articles/beware-the-kolmogorov-smirnov-test.}.
\item We repeat the above steps and generate $10^{4}$ mock GRB samples with different model parameter sets to determine the best parameters and their confidence level. In case of $P_{\rm K-S}<10^{-4}$, the hypothesis that two data sets are from the same parent can be rejected at the 3 $\sigma$ confidence level. We therefore estimate the confidence level with $P_{\rm K-S}>10^{-4}$.
    \end{itemize}

Figure \ref{MCdata} shows $P_{K-s}$ as a function of $L_{b, 0}$, $\alpha_{1}$, $\alpha_{2}$, $\epsilon$ and $k_{\rm b}$. We illustrate the comparison between observed GRB sample and the simulated GRB sample generated with the best parameter set in the $\log L_{\rm p}-\log(1+z)$ plane in Figure 5. The one-dimensional distributions of $L_{\rm p}$ and $z$, together with their $P_{\rm KS}$ values, are also shown in Figure 5. One can observe that the best parameter set can well reproduce the observed sample with our simulations. Constraints on the parameters  with the current sample are $ L_{\rm b, 0}/10^{50} {\rm erg s^{-1}}=17.2^{+36.8}_{-10.3}$, $\alpha_{1}=1.02^{+0.35}_{-0.56}$, $\alpha_{2}=2.51^{+1.89}_{-0.56}$,
$\epsilon=0.35^{+0.44}_{-0.19}$ and $k_{\rm b}=1.14^{+0.99}_{-0.47}$. The derived $L_{b, 0}$, $\alpha_{1}$, and $\alpha_{2}$ are generally consistent with that reported in previous paper within error bars (e.g., Schmidt 2001; Stern et al. 2002; Guetta et al.
2005; Dai \&Zhang 2005; Daigne et al. 2006; Liang et al. 2007). The $\epsilon$ value varies from 0.2 to 0.6 reported in previous paper (Li 2008; Modjaz et al. 2008; Qin et al. 2010). The derived $\epsilon$ value in this analysis is consistent with previous results and also agrees with theoretical expectation of low metallicity for LGRB (e.g., Wolf \& Podsiadlowski 2007). The derived $k_b$ value from our simulation analysis is smaller than that derived from the $\tau$ statistics method, but it has a very large error bar.

\section{Effects of Observational Biases on the $\tau$ Statistics method}
As mentioned in \S 1, the current sample for our analysis is not a completed sample in redshift for a given instrument threshold. It is suffered biases on the flux truncation effect, trigger probability, and redshift measurement. The true instrument threshold for a GRB trigger is complicated. For example, the BAT threshold is generally taken as $1\times 10^{-8}$ ergs cm$^{-2}$ s$^{-1}$ as reported by the BAT team, but the trigger probability of a GRB with a peak flux close to the instrument threshold is much lower than that of high-flux GRBs (e.g., Stern et al. 2001 Qin et al. 2010). In addition, in the image trigger mode, a GRB trigger also depends on the burst duration, hence the burst fluence (Band 2006, Sakamoto et al. 2007, Virgili et al. 2009). The biases of redshift measurement are even much complicated. As reported by Qin et al. (2010), the redshift measure probability of a bright GRB tends to higher than a dim burst, but there are several other detector and observational biases that affect the measurement of the redshift and are independent of the brightness of a GRB, such as galactic dust extinction, redshift desert, host galaxy extinction, etc. (e.g., Coward et al. 2013). These factors complicate the completeness of the sample selection. One should keep in mind that the non-parametric $\tau$-statistics method is employed for estimating the intrinsic correlation in a complete sample at a given selection threshold.

To examine the threshold truncation effect on the result of the non-parametric $\tau$-statistics method, we plot the $k$ value as a function of BAT threshold for our sample in Figure 6. One can observe that $k$ is sensitive to $F_{\rm th}$. A larger $k$ value is obtained by using a lower $F_{\rm th}$, and the upper limit of $k$ is that derived from the best linear fit to the data without considering the data truncation effect, as we mark also in Figure 6. Adopting the BAT threshold as $1\times 10^{-8}$ erg cm$^{-2}$ s$^{-1}$, we get $k=1.49\pm0.19$ accordingly. Note that the current sample is not a complete sample for this threshold since it is suffered biases of GRB trigger and redshift measurement. These biases would make a gap between the data and the threshold line since GRBs with a flux close to the threshold tend to have a low trigger probability and redshift measurement probability. This effect leads to overestimate the $k$ value. We investigate how these biases affect the estimate of $k$ with the $\tau$-statistics method based on our simulation analysis. We model the these biases by fitting the GRB distribution in the $\log L-\log (1+z)$ plane in our simulation analysis. With the derived model parameters reported in \S 4, we reproduce the a mock sample and calculate $k$ as a function of $F_{\rm th}$, which is also shown in Figure 6. We have $k=1.66\pm 0.13$ by taking $F_{\rm th}=1\times 10^{-8}$ erg cm$^{-2}$ s$^{-1}$. It is roughly consistent with that derived from the observed GRB sample, but is much larger than the input of our model for reproducing the sample ($k_b=1.14$). We further generate mock {\em complete} BAT samples of 258 and 1000 GRBs with the best model parameters, says, all GRBs with a flux over the threshold line $F_{\rm th, BAT}=1\times 10^{-8}$ erg cm$^{-2}$ s$^{-1}$ are included in the sample without considering the trigger and redshift measurement probabilities. $k$ as a function of $F_{\rm th}$ is also shown in Figure 6. We have $k=0.94\pm 0.14$ and $k=0.80\pm 0.09$ from the two mock samples with $F_{\rm th}=F_{\rm th, BAT}$, respectively. Note that the two mock {\em complete} samples are generated with a threshold of $F_{\rm th, BAT}$. They are ``{\em complete}'' only for $F_{\rm th}\geq F_{\rm th, BAT}$. Therefore, the $k$ value derived from the mock {\em complete} BAT samples significantly depends on the threshold if the adopted $F_{\rm th}$ is less than $F_{\rm th, BAT}$. A sub-sample in different size selected from the mock {\em complete} BAT samples by adopting a $F_{\rm th}$ being greater $F_{\rm th, BAT}$ is equivalent to a {\em complete} sample for the given threshold, and the derived $k$ then weakly depends on $F_{\rm th}$, as shown in Fig.6. To further investigate this issue, we generate mock {\em complete} samples of 1000 GRBs with different thresholds and show also the derived $k$ from these samples as a function of the corresponding threshold in Figure 6. One can see that $k$ does not depend on $F_{\rm th}$, and its mean is $0.75\pm 0.09$.

\section{Conclusions and Discussion}
We have revisited the intrinsic $L_{\rm p}-z$ relation with two different approaches. Firstly, by simplifying the luminosity of individual GRBs evolves as $L_{p}\propto{\rm }(1+z)^{k}$, we get $k=1.49\pm0.19$ with the non-parameterized $\tau$ statistics method. Secondly, by characterizing the luminosity function of long GRBs as a smoothly broken power-law with $L_{\rm b}$ evolving as $L_{\rm b}\propto (1+z)^{k_{\rm b}}$ and parameterizing the trigger probability and redshift measurement probability as a function of the observed peak flux, we get $k_{\rm b}=1.14^{+0.99}_{-0.47}$ from simulations under the assumption that the long GRB rate follows the star formation rate incorporating with cosmic metallicity history. Further more, by removing the observational biases of the GRB trigger and redshift measurement based on our simulation analysis results, we generate mock {\em complete} samples of 258 and 1000 GRBs based on our simulation analysis and obtain $k=0.94\pm 0.14$ and $k=0.80\pm 0.09$ with the $\tau$ statistics method, indicating that these observational biases may lead to overestimate the $k$ value with this method.

 Besides the observational biases of the GRB trigger and redshift measurement, another issue regarding the completeness of the sample is contamination of different GRB populations. Looking at Figure 1, one can observe that most of the GRBs at $z < 0.5$ tightly track the flux sensitivity of the BAT, and there is an obvious deficit of medium to high-luminosity GRBs at low redshift. Note that low luminosity GRBs may be a distinct population from typical GRBs and their event rate is much higher than typical GRBs (e.g., Liang et al. 2007). The detection rate of low luminosity at low-z then should be much higher than high luminosity GRBs. This may result in most GRBs at $z < 1$ tightly track the flux sensitivity of the BAT. The deficit of medium to high-luminosity GRBs at low redshift (such as GRB 030329) may be due to their low event rate and a small observable volume at low redshift. Although we have removed some typical low-luminosity GRBs (such as GRB 980425 and 060218), we still cannot convincingly remove the contamination of this population at low redshift. In addition, GRBs from compact star Mergers are also usually detected at low redshift, but they usually show up an spiky pulse with bright peak luminosity (e.g., L\"{u} et al. 2014). We exclude those GRBs at $z<0.5$ and derive an apparent relation of $\log L_{\rm p}=(50.13\pm 0.10)+(2.66\pm 0.18)\log[(1+z)]$ the with the ordinary least squares algorithm. The $\tau$ statistics gives $k=1.51\pm 0.20$. The slope of the apparent $\log L-\log (1+z)$ relation is slightly smaller than that derived from the low-z GRB included sample, and the slope estimated with the $\tau$ statistics method is consistent with that from the global sample. Therefore, the possible contamination of low luminosity GRBs and GRBs from compact star mergers is not significantly affects the $\tau$ statistics results.

The narrowness of the BAT energy band may also have influence on our results. Note that a GRB spectrum is usually fitted with the Band function. Its $E_p$ value may vary from several keVs to GeVs (e.g., Zhang et al. 2011). Being due to the narrowness of the BAT bandpass (15-150 keV), the observed fluxes are only a slice of the GRB spectra. This makes uncertainty of measuring the bolometric luminosity of GRBs. In order to avoid possible unreasonable extrapolation without broadband spectral information and to keep consistency with the BAT threshold in the 15-150 keV, we calculate the luminosity with the observed fluxes only. Although the luminosity is not the bolometric one and is in a different rest-frame band for each GRB, using the observed flux and corresponding instrument threshold in the same bandpass for our statistical purpose would not lead to significant bias in our results. To justify this argument, we also calculate the luminosity in the 15-150 KeV in the rest frame for the GRBs in our sample and derive the $k$ value accordingly. We get $k=1.52\pm0.20$, which is comparable to the result derived from the sample that GRB luminosity is calculated in the observed 15-150 keV band, as reported above. Note that the observed $\log L-\log (1+z)$ relation has a large intrinsic scatter. Systematical uncertainty of the GRB luminosity with the BAT observations in a narrow energy band may be contributed to the intrinsic scatter in some extend. From Eq. (3), one can observe that the result of the $\tau$ statistics is not significantly affected by the intrinsic scatter.

Cosmic evolution of GRB luminosity has been extensively studied. As mentioned in \S 1, samples used in previous papers have no
redshift information available (e.g., Lloyd-Ronning et al. 2002; Yonetoku et al. 2004; Kocevski \& Liang 2006; Tan et al.2013) or have a small size or are collected from GRB surveys with instruments in different energy bands (e.g., Salvaterra et al. 2012; Petrosian et al. 2009). Before the {\em Swift} mission era, the large and uniform sample observed with the Burst and Transient Source Experiment (BATSE) on board Compton Gamma-Ray Observatory (CGRO) is used for analysis. Being lack of redshift information, the redshift and luminosity of BATSE GRBs were estimated with some empirical luminosity-indicator relations. For example, by estimating the redshifts and luminosities with the luminosity-variability relationship (Fenimore \& Ramirez-Ruiz 2000) for 220 BATSE GRBs, Lloyd-Ronning et al. (2002) obtained $k=1.4\pm0.5$. By using the relation between luminosity and the peak energy of the $\nu f_\nu$ spectrum for a sample of 689 BATSE GRBs, Yonetoku et al. (2004) suggested $k=2.6\pm0.2$. Estimating the redshifts with the spectral lag-luminosity correlation (Norris et al. 2000) for $\sim 900$ BATSE GRBs, Kocevski \& Liang (2006) reported $k=1.7\pm0.3$. With a sample of 86 GRBs that have spectroscopic redshift measurement, Petrosian et al. (2009) reported $k=[1.75,2.74]$. Measuring the $\log L-\log (1+z)$ distribution by assuming that the GRB rate follows the SFR, Salvaterra et al. (2012) reported $k_{\rm b}=2.3\pm0.6$ or strong comic evolution in GRB number density with a selected redshift-known sample of 52 Swift/BAT GRBs. Employing the $L-E_{\rm p}$ relation to estimate the redshifts for a sample of 498 BAT GRBs, Tan et al. (2013) obtained $k_b\sim 2$. The derived $k$ and $k_b$ values from our analysis are systematically smaller than previous results by these authors. The discrepancy may be due to the sample selections. The large flux-truncated {\em CGRO/BATSE} GRB sample and {\em Swift}/BAT sample were usually used (e.g., Lloyd-Ronning et al. 2002; Yonetoku et al. 2004; Kocevski \& Liang 2006; Tan et al.2013), but they have no spectroscopic redshift information available. Pseudo redhsifts estimated with the empirical luminosity-indicator relations have great uncertainty. Although the samples used in some previous papers have redshift information, they are rather small and even collected from GRBs detected by instruments in different energy bands (e.g., Salvaterra et al. 2012; Petrosian et al. 2009). For example, the GRB sample used in Petrosian et al. (2009) is not selected by a given threshold since they were detected by instruments in different thresholds and energy bands, i.e., {\em BATSE} (25-2000 keV), {\em Beppo-SAX} (40-700 keV) , {\em HETE-2} (2-400 keV), and {\em Swift} (15-150 keV). More importantly, the observational biases of GRB trigger and redshift measurement are not taken into account in previous analysis. This may lead to over-estimate the $k$ value, as we mentioned above. Qin et al. (2010) showed that the Swift observations can be well reproduced by assuming that the GRB rate follows the SFR incorporating the cosmic metallicity history and the cosmic evolution in GRB number density without introduce any cosmic luminosity evolution (see also Wanderman \& Prian 2010). Coward et al. (2013) even proposed that it is not necessary to invoke luminosity evolution with redshift to explain the observed GRB rate at high-z by carefully taking selection effects into account (see also Howell \& Coward, 2013). Thanks to {\em Swift}/BAT, we now have a considerable uniform sample of 258 GRBs with redshift measurement. However, with the large uncertain of $k_b$ derived from our simulation analysis, we still cannot convincingly argue a robust evolution feature of GRB luminosity.
\begin{acknowledgements}
We acknowledge the use of the public data from the Swift data
archive. We thank of the We thank the anonymous referee, Bing Zhang, Zi-Gao Dai, and
Shuang-Nan Zhang for helpful suggestion. This work is supported by the National Basic Research
Program (973 Programme) of China (Grant No. 2014CB845800),
the National Natural Science Foundation of China (Grant Nos. 11533003, 11363002,11303005, U1331202, 11322328, 11373036), the Guangxi Science
Foundation (2013GXNSFFA019001,2014GXNSFAA118011),the One-Hundred-Talents Program, the Youth Innovation Promotion Association, the Strategic Priority
Research Program ``The Emergence of Cosmological
Structures''(Grant No. XDB09000000), and the Key Laboratory for the Structure and
Evolution of Celestial Objects, of the Chinese Academy of Sciences.
\end{acknowledgements}

\clearpage
\begin{deluxetable}{lllll|lllll}
\rotate
\tablewidth{550pt}
\tabletypesize{\footnotesize}
\tablecaption{The data of our GRB sample}
\tablehead{\colhead{GRB}&\colhead{$z$}&\colhead{$\log_{10}L_{\rm p}$$^{a}$}&\colhead{$P^{b}$}&\colhead{$\Gamma^{c}$}&\colhead{GRB}&\colhead{$z$}&\colhead{$\log_{10}L_{\rm p}$$^{a}$}&\colhead{$P^{b}$}&\colhead{$\Gamma^{c}$} \\  \colhead{}& \colhead{}&\colhead{(erg/s)}&\colhead{(ph/cm$^{2}$/s)}&\colhead{}& \colhead{}&\colhead{}&\colhead{(erg/s)} &\colhead{(ph/cm$^{2}$/s)} }

\startdata

050126	&	1.29	&	50.50	$\pm$	0.12	&	0.7	$\pm$	0.2	&	1.34	$\pm$	0.15	&	050915A	&	 2.5273	&	51.15	$\pm$	 0.10	 &	 0.8	$\pm$	0.1	&	1.39	$\pm$	 0.17	\\
050223	&	0.5915	&	49.77	$\pm$	0.11	&	0.7	$\pm$	0.2	&	1.85	$\pm$	0.17	&	050922C	&	 2.198	&	52.00	$\pm$	 0.03	 &	 7.26	$\pm$	0.32	&	1.37	 $\pm$	0.06	 \\
050315	&	1.949	&	51.53	$\pm$	0.06	&	1.9	$\pm$	0.2	&	2.11	$\pm$	0.09	&	051001	&	 2.4296	&	51.15	$\pm$	 0.12	 &	 0.5	$\pm$	0.1	&	2.05	$\pm$	 0.15	\\
050318	&	1.44	&	51.36	$\pm$	0.03	&	3.2	$\pm$	0.2	&	1.90	$\pm$	0.10	&	051006	&	 1.059	&	50.70	$\pm$	 0.09	 &	 1.6	$\pm$	0.3	&	1.51	$\pm$	 0.17	\\
050319	&	3.24	&	51.94	$\pm$	0.11	&	1.5	$\pm$	0.2	&	2.02	$\pm$	0.19	&	051016B	&	 0.9364	&	50.59	$\pm$	 0.06	 &	 1.3	$\pm$	0.2	&	2.40	$\pm$	 0.23	\\
050401	&	2.9	&	52.41	$\pm$	0.05	&	10.7	$\pm$	0.9	&	1.40	$\pm$	0.07	&	051109A	&	 2.346	&	51.83	$\pm$	 0.11	 &	 3.9	$\pm$	0.7	&	1.51	$\pm$	 0.20	\\
050406	&	2.44	&	51.16	$\pm$	0.18	&	0.36	$\pm$	0.10	&	2.43	$\pm$	0.35	&	 051111	&	1.549	&	51.24	 $\pm$	 0.04	&	2.7	$\pm$	0.2	&	1.32	 $\pm$	0.06	 \\
050416A	&	0.6535	&	50.81	$\pm$	0.05	&	4.88	$\pm$	0.48	&	3.08	$\pm$	0.22	&	 051117B	&	0.481	&	49.41	 $\pm$	 0.15	&	0.5	$\pm$	0.1	&	1.53	 $\pm$	0.31	 \\
050505	&	4.27	&	51.98	$\pm$	0.10	&	1.9	$\pm$	0.3	&	1.41	$\pm$	0.12	&	060108	&	 2.03	&	51.15	$\pm$	 0.09	 &	 0.8	$\pm$	0.1	&	2.03	$\pm$	 0.17	\\
050730	&	3.96855	&	51.45	$\pm$	0.13	&	0.6	$\pm$	0.1	&	1.53	$\pm$	0.11	&	060115	&	 3.53	&	51.66	$\pm$	 0.08	 &	 0.9	$\pm$	0.1	&	1.76	$\pm$	 0.12	\\
050802	&	1.71	&	51.39	$\pm$	0.08	&	2.8	$\pm$	0.4	&	1.54	$\pm$	0.13	&	060124	&	 2.3	&	51.28	$\pm$	 0.11	 &	 0.9	$\pm$	0.2	&	1.84	$\pm$	 0.19	\\
050814	&	5.3	&	51.98	$\pm$	0.17	&	0.7	$\pm$	0.3	&	1.80	$\pm$	0.17	&	060206	&	 4.045	&	52.26	$\pm$	0.05	 &	 2.79	$\pm$	0.17	&	1.69	$\pm$	 0.08	\\
050819	&	2.5043	&	51.33	$\pm$	0.21	&	0.4	$\pm$	0.1	&	2.71	$\pm$	0.29	&	060210	&	 3.91	&	52.14	$\pm$	 0.06	 &	 2.7	$\pm$	0.3	&	1.53	$\pm$	 0.09	\\
050820A	&	2.612	&	51.63	$\pm$	0.06	&	2.5	$\pm$	0.2	&	1.25	$\pm$	0.12	&	060223A	&	 4.41	&	52.05	$\pm$	 0.09	 &	 1.4	$\pm$	0.2	&	1.74	$\pm$	 0.12	\\
050824	&	0.83	&	50.07	$\pm$	0.16	&	0.5	$\pm$	0.2	&	2.76	$\pm$	0.38	&	060418	&	 1.49	&	51.67	$\pm$	 0.03	 &	 6.5	$\pm$	0.4	&	1.70	$\pm$	 0.06	\\
050826	&	0.297	&	48.87	$\pm$	0.14	&	0.4	$\pm$	0.1	&	1.16	$\pm$	0.31	&	060505	&	 0.089	&	48.60	$\pm$	 0.12	 &	 2.7	$\pm$	0.6	&	1.29	$\pm$	 0.28	\\
050904	&	6	&	51.67	$\pm$	0.14	&	0.6	$\pm$	0.2	&	1.25	$\pm$	0.07	&	060510B	&	4.9	 &	51.80	$\pm$	0.10	&	 0.6	 $\pm$	0.1	&	1.78	$\pm$	0.08	 \\
050908	&	3.35	&	51.57	$\pm$	0.13	&	0.7	$\pm$	0.1	&	1.88	$\pm$	0.17	&	060512	&	 0.4428	&	49.57	$\pm$	 0.10	 &	 0.9	$\pm$	0.2	&	2.48	$\pm$	 0.30	\\
																															
060522	&	5.11	&	51.69	$\pm$	0.16	&	0.6	$\pm$	0.2	&	1.56	$\pm$	0.15	&	061110A	&	 0.758	&	49.90	$\pm$	 0.10	 &	 0.5	$\pm$	0.1	&	1.67	$\pm$	 0.12	\\
060526	&	3.21	&	51.96	$\pm$	0.12	&	1.7	$\pm$	0.2	&	2.01	$\pm$	0.24	&	061110B	&	 3.44	&	51.01	$\pm$	 0.14	 &	 0.5	$\pm$	0.1	&	1.03	$\pm$	 0.16	\\
060602A	&	0.787	&	49.94	$\pm$	0.18	&	0.6	$\pm$	0.2	&	1.25	$\pm$	0.16	&	061121	&	 1.314	&	52.01	$\pm$	 0.01	 &	 21.1	$\pm$	0.5	&	1.41	$\pm$	 0.03	\\
060604	&	2.1357	&	50.84	$\pm$	0.33	&	0.3	$\pm$	0.1	&	2.01	$\pm$	0.42	&	061210	&	 0.41	&	50.29	$\pm$	 0.04	 &	 5.3	$\pm$	0.5	&	1.56	$\pm$	 0.28	\\
060605	&	3.8	&	51.35	$\pm$	0.16	&	0.5	$\pm$	0.1	&	1.55	$\pm$	0.20	&	061222A	&	 2.088	&	52.02	$\pm$	0.02	 &	 8.5	$\pm$	0.3	&	1.35	$\pm$	 0.04	\\
060607A	&	3.082	&	51.61	$\pm$	0.05	&	1.4	$\pm$	0.1	&	1.47	$\pm$	0.08	&	061222B	&	 3.355	&	51.97	$\pm$	 0.12	 &	 1.6	$\pm$	0.4	&	1.97	$\pm$	 0.13	\\
060707	&	3.43	&	51.66	$\pm$	0.12	&	1.0	$\pm$	0.2	&	1.68	$\pm$	0.13	&	070103	&	 2.6208	&	51.52	$\pm$	 0.10	 &	 1.0	$\pm$	0.2	&	1.95	$\pm$	 0.21	\\
060714	&	2.71	&	51.64	$\pm$	0.06	&	1.3	$\pm$	0.1	&	1.93	$\pm$	0.11	&	070110	&	 2.352	&	51.04	$\pm$	 0.10	 &	 0.6	$\pm$	0.1	&	1.58	$\pm$	 0.12	\\
060719	&	1.532	&	51.27	$\pm$	0.05	&	2.2	$\pm$	0.2	&	1.91	$\pm$	0.11	&	070129	&	 2.3384	&	51.15	$\pm$	 0.11	 &	 0.6	$\pm$	0.1	&	2.01	$\pm$	 0.15	\\
060729	&	0.54	&	49.91	$\pm$	0.05	&	1.2	$\pm$	0.1	&	1.75	$\pm$	0.14	&	070208	&	 1.165	&	50.60	$\pm$	 0.13	 &	 0.9	$\pm$	0.2	&	1.94	$\pm$	 0.36	\\
060814	&	0.84	&	51.13	$\pm$	0.02	&	7.3	$\pm$	0.3	&	1.53	$\pm$	0.03	&	070318	&	 0.836	&	50.50	$\pm$	 0.04	 &	 1.8	$\pm$	0.2	&	1.42	$\pm$	 0.08	\\
060904B	&	0.703	&	50.49	$\pm$	0.04	&	2.4	$\pm$	0.2	&	1.64	$\pm$	0.14	&	070411	&	 2.954	&	51.49	$\pm$	 0.07	 &	 0.9	$\pm$	0.1	&	1.72	$\pm$	 0.10	\\
060906	&	3.685	&	52.19	$\pm$	0.09	&	2.0	$\pm$	0.3	&	2.03	$\pm$	0.11	&	070419A	&	 0.97	&	49.81	$\pm$	 0.28	 &	 0.2	$\pm$	0.1	&	2.35	$\pm$	 0.25	\\
060908	&	1.8836	&	51.49	$\pm$	0.03	&	3.2	$\pm$	0.2	&	1.33	$\pm$	0.07	&	070506	&	 2.31	&	51.28	$\pm$	 0.08	 &	 0.96	$\pm$	0.13	&	1.73	 $\pm$	0.17	 \\
060926	&	3.208	&	52.02	$\pm$	0.13	&	1.09	$\pm$	0.14	&	2.54	$\pm$	0.23	&	 070508	&	0.82	&	51.62	 $\pm$	 0.01	&	24.1	$\pm$	0.6	&	1.36	 $\pm$	 0.03	\\
060927	&	5.6	&	52.52	$\pm$	0.06	&	2.7	$\pm$	0.2	&	1.65	$\pm$	0.08	&	070521	&	 0.553	&	50.67	$\pm$	0.02	 &	 6.5	$\pm$	0.3	&	1.38	$\pm$	 0.04	\\
061007	&	1.261	&	51.74	$\pm$	0.01	&	14.6	$\pm$	0.4	&	1.03	$\pm$	0.03	&	070529	 &	2.4996	&	51.39	$\pm$	 0.14	 &	1.4	$\pm$	0.4	&	1.34	$\pm$	 0.16	\\
061021	&	0.3463	&	50.21	$\pm$	0.02	&	6.1	$\pm$	0.3	&	1.30	$\pm$	0.06	&	070611	&	 2.04	&	51.07	$\pm$	 0.14	 &	 0.8	$\pm$	0.2	&	1.66	$\pm$	 0.22	\\
																															
070612A	&	0.617	&	50.15	$\pm$	0.12	&	1.5	$\pm$	0.4	&	1.69	$\pm$	0.10	&	080411	&	 1.03	&	52.13	$\pm$	 0.01	 &	 43.2	$\pm$	0.9	&	1.75	$\pm$	 0.03	\\
070714B	&	0.92	&	50.77	$\pm$	0.04	&	2.7	$\pm$	0.2	&	1.36	$\pm$	0.19	&	080413A	&	 2.433	&	52.04	$\pm$	 0.03	 &	 5.6	$\pm$	0.2	&	1.57	$\pm$	 0.06	\\
070802	&	2.45	&	50.98	$\pm$	0.15	&	0.4	$\pm$	0.1	&	1.79	$\pm$	0.27	&	080413B	&	 1.1	&	51.84	$\pm$	 0.02	 &	 18.7	$\pm$	0.8	&	1.80	$\pm$	 0.06	\\
071010A	&	2.17	&	51.62	$\pm$	0.06	&	1.9	$\pm$	0.2	&	2.04	$\pm$	0.14	&	080430	&	 0.767	&	50.61	$\pm$	 0.03	 &	 2.6	$\pm$	0.2	&	1.73	$\pm$	 0.09	\\
071010B	&	1.1	&	51.31	$\pm$	0.03	&	6.3	$\pm$	0.4	&	1.36	$\pm$	0.07	&	080516	&	3.2	 &	51.91	$\pm$	0.14	&	 1.8	 $\pm$	0.3	&	1.82	$\pm$	0.27	 \\
071020	&	0.98	&	50.42	$\pm$	0.17	&	0.8	$\pm$	0.3	&	2.33	$\pm$	0.37	&	080520	&	 1.545	&	50.89	$\pm$	 0.18	 &	 0.5	$\pm$	0.1	&	2.90	$\pm$	 0.51	\\
071021	&	2.452	&	51.19	$\pm$	0.10	&	0.7	$\pm$	0.1	&	1.70	$\pm$	0.21	&	080603B	&	 2.69	&	52.01	$\pm$	 0.04	 &	 3.5	$\pm$	0.2	&	1.78	$\pm$	 0.07	\\
071028B	&	0.94	&	50.52	$\pm$	0.25	&	1.4	$\pm$	0.5	&	1.45	$\pm$	0.25	&	080604	&	 1.416	&	50.43	$\pm$	 0.13	 &	 0.4	$\pm$	0.1	&	1.78	$\pm$	 0.18	\\
071031	&	2.692	&	51.42	$\pm$	0.16	&	0.5	$\pm$	0.1	&	2.42	$\pm$	0.29	&	080605	&	 1.6398	&	52.17	$\pm$	 0.02	 &	 19.9	$\pm$	0.6	&	1.36	$\pm$	 0.03	\\
071112C	&	0.823	&	51.12	$\pm$	0.06	&	8.0	$\pm$	1.0	&	1.09	$\pm$	0.07	&	080607	&	 3.036	&	52.75	$\pm$	 0.03	 &	 23.1	$\pm$	1.1	&	1.31	$\pm$	 0.04	\\
071117	&	1.331	&	51.78	$\pm$	0.02	&	11.3	$\pm$	0.4	&	1.57	$\pm$	0.06	&	080707	 &	1.23	&	50.68	$\pm$	 0.05	 &	1.0	$\pm$	0.1	&	1.77	$\pm$	 0.19	\\
071122	&	1.14	&	50.20	$\pm$	0.29	&	0.4	$\pm$	0.2	&	1.77	$\pm$	0.31	&	080710	&	 0.845	&	50.27	$\pm$	 0.09	 &	 1.0	$\pm$	0.2	&	1.47	$\pm$	 0.23	\\
080207	&	2.0858	&	51.15	$\pm$	0.15	&	1.0	$\pm$	0.3	&	1.58	$\pm$	0.06	&	080721	&	 2.602	&	52.50	$\pm$	 0.05	 &	 20.9	$\pm$	1.8	&	1.11	$\pm$	 0.08	\\
080210	&	2.641	&	51.65	$\pm$	0.07	&	1.6	$\pm$	0.2	&	1.77	$\pm$	0.12	&	080804	&	 2.2045	&	51.57	$\pm$	 0.06	 &	 3.1	$\pm$	0.4	&	1.19	$\pm$	 0.09	\\
080310	&	2.4266	&	51.67	$\pm$	0.09	&	1.3	$\pm$	0.2	&	2.32	$\pm$	0.16	&	080805	&	 1.505	&	50.87	$\pm$	 0.04	 &	 1.1	$\pm$	0.1	&	1.53	$\pm$	 0.07	\\
080319B	&	0.937	&	51.72	$\pm$	0.01	&	24.8	$\pm$	0.5	&	1.04	$\pm$	0.02	&	080810	 &	3.35	&	51.78	$\pm$	 0.05	 &	2.0	$\pm$	0.2	&	1.34	$\pm$	 0.06	\\
080319C	&	1.95	&	51.75	$\pm$	0.03	&	5.2	$\pm$	0.3	&	1.37	$\pm$	0.07	&	080905B	&	 2.374	&	51.04	$\pm$	 0.11	 &	 0.5	$\pm$	0.1	&	1.78	$\pm$	 0.15	\\
080330	&	1.51	&	51.02	$\pm$	0.15	&	0.9	$\pm$	0.2	&	2.53	$\pm$	0.45	&	080906	&	 2	&	51.12	$\pm$	0.09	 &	 1.0	$\pm$	0.2	&	1.59	$\pm$	 0.09	\\
																															
080913	&	6.44	&	52.15	$\pm$	0.12	&	1.40	$\pm$	0.20	&	1.36	$\pm$	0.15	&	 090423	&	8.2	&	52.84	$\pm$	 0.09	 &	1.7	$\pm$	0.2	&	1.48	$\pm$	 0.07	\\
080916A	&	0.689	&	50.51	$\pm$	0.03	&	2.7	$\pm$	0.2	&	1.36	$\pm$	0.15	&	090424	&	 0.544	&	51.70	$\pm$	 0.01	 &	 71.0	$\pm$	2.0	&	1.90	$\pm$	 0.10	\\
080928	&	1.692	&	51.32	$\pm$	0.04	&	2.1	$\pm$	0.1	&	1.63	$\pm$	0.05	&	090426	&	 2.609	&	51.87	$\pm$	 0.10	 &	 2.4	$\pm$	0.3	&	1.93	$\pm$	 0.22	\\
081007	&	0.5295	&	50.24	$\pm$	0.07	&	2.6	$\pm$	0.4	&	1.77	$\pm$	0.12	&	090429B	&	 9.4	&	53.00	$\pm$	 0.13	 &	 1.6	$\pm$	0.2	&	1.96	$\pm$	 0.13	\\
081008	&	1.9685	&	51.24	$\pm$	0.04	&	1.3	$\pm$	0.1	&	2.51	$\pm$	0.20	&	090516A	&	 4.109	&	52.11	$\pm$	 0.08	 &	 1.6	$\pm$	0.2	&	1.62	$\pm$	 0.03	\\
081028A	&	3.038	&	51.06	$\pm$	0.18	&	0.5	$\pm$	0.1	&	1.69	$\pm$	0.07	&	090519	&	 3.9	&	51.22	$\pm$	 0.20	 &	 0.6	$\pm$	0.2	&	1.84	$\pm$	 0.11	\\
081029	&	3.8479	&	51.34	$\pm$	0.40	&	0.5	$\pm$	0.2	&	1.25	$\pm$	0.38	&	090618	&	 0.54	&	51.43	$\pm$	 0.01	 &	 38.9	$\pm$	0.8	&	1.02	$\pm$	 0.20	\\
081118	&	2.58	&	51.32	$\pm$	0.18	&	0.6	$\pm$	0.2	&	1.43	$\pm$	0.18	&	090715B	&	 3	&	52.06	$\pm$	0.04	 &	 3.8	$\pm$	0.2	&	1.72	$\pm$	 0.02	\\
081121	&	2.512	&	51.84	$\pm$	0.11	&	4.4	$\pm$	1.0	&	2.10	$\pm$	0.16	&	090726	&	 2.71	&	51.50	$\pm$	 0.17	 &	 0.7	$\pm$	0.2	&	1.57	$\pm$	 0.07	\\
081203A	&	2.1	&	51.61	$\pm$	0.04	&	2.9	$\pm$	0.2	&	1.21	$\pm$	0.12	&	090809	&	 2.737	&	51.35	$\pm$	0.13	 &	 1.1	$\pm$	0.2	&	1.34	$\pm$	 0.24	\\
081221	&	2.26	&	52.58	$\pm$	0.01	&	18.2	$\pm$	0.5	&	1.54	$\pm$	0.06	&	090812	 &	2.452	&	51.74	$\pm$	 0.03	 &	3.6	$\pm$	0.2	&	2.25	$\pm$	 0.19	\\
081222	&	2.77	&	52.26	$\pm$	0.02	&	7.7	$\pm$	0.2	&	1.86	$\pm$	0.0	&	090814A	&	 0.696	&	49.88	$\pm$	0.18	 &	 0.6	$\pm$	0.2	&	1.24	$\pm$	 0.05	\\
090102	&	1.547	&	51.56	$\pm$	0.07	&	5.5	$\pm$	0.8	&	1.48	$\pm$	0.03	&	090913	&	 6.7	&	52.18	$\pm$	 0.12	 &	 1.4	$\pm$	0.2	&	1.81	$\pm$	 0.19	\\
090113	&	1.7493	&	51.39	$\pm$	0.04	&	2.5	$\pm$	0.2	&	1.36	$\pm$	0.08	&	090926B	&	 1.24	&	51.15	$\pm$	 0.04	 &	 3.2	$\pm$	0.3	&	1.54	$\pm$	 0.0	\\
090205	&	4.7	&	51.91	$\pm$	0.16	&	0.5	$\pm$	0.1	&	2.15	$\pm$	0.23	&	090927	&	 1.37	&	51.09	$\pm$	0.06	 &	 2.0	$\pm$	0.2	&	1.80	$\pm$	 0.20	\\
090313	&	3.375	&	51.65	$\pm$	0.32	&	0.8	$\pm$	0.3	&	1.60	$\pm$	0.10	&	091018	&	 0.971	&	51.51	$\pm$	 0.02	 &	 10.3	$\pm$	0.4	&	2.30	$\pm$	 0.06	\\
090407	&	1.4485	&	50.61	$\pm$	0.09	&	0.6	$\pm$	0.1	&	1.91	$\pm$	0.29	&	091020	&	 1.71	&	51.58	$\pm$	 0.04	 &	 4.2	$\pm$	0.3	&	1.53	$\pm$	 0.07	\\
090418A	&	1.608	&	51.16	$\pm$	0.08	&	1.9	$\pm$	0.3	&	1.73	$\pm$	0.29	&	091024	&	 1.092	&	50.78	$\pm$	 0.07	 &	 2.0	$\pm$	0.3	&	1.20	$\pm$	 0.08	\\
																															
091029	&	2.752	&	51.78	$\pm$	0.03	&	1.8	$\pm$	0.1	&	1.88	$\pm$	0.06	&	100816A	&	 0.8034	&	51.24	$\pm$	 0.02	 &	 10.90	$\pm$	0.40	&	1.16	 $\pm$	0.06	 \\
091109A	&	3.5	&	51.62	$\pm$	0.19	&	1.3	$\pm$	0.4	&	1.31	$\pm$	0.25	&	100901A	&	 1.408	&	50.67	$\pm$	0.13	 &	 0.8	$\pm$	0.2	&	1.52	$\pm$	 0.21	\\
091127	&	0.49	&	51.40	$\pm$	0.03	&	46.5	$\pm$	2.7	&	2.05	$\pm$	0.07	&	100902A	 &	4.5	&	52.07	$\pm$	 0.08	 &	 1.0	$\pm$	0.1	&	1.98	$\pm$	 0.13	\\
091208B	&	1.063	&	51.71	$\pm$	0.03	&	15.2	$\pm$	1.0	&	1.74	$\pm$	0.11	&	100906A	 &	1.727	&	52.03	$\pm$	 0.02	 &	10.1	$\pm$	0.4	&	1.78	 $\pm$	0.03	 \\
100219A	&	4.7	&	51.35	$\pm$	0.19	&	0.4	$\pm$	0.1	&	1.34	$\pm$	0.25	&	101219B	&	 0.5519	&	49.64	$\pm$	0.29	 &	 0.6	$\pm$	0.3	&	1.56	$\pm$	 0.16	\\
100302A	&	4.813	&	51.69	$\pm$	0.15	&	0.5	$\pm$	0.1	&	1.72	$\pm$	0.19	&	110128A	&	 2.339	&	51.07	$\pm$	 0.16	 &	 0.8	$\pm$	0.2	&	1.31	$\pm$	 0.30	\\
100316B	&	1.18	&	50.82	$\pm$	0.04	&	1.30	$\pm$	0.10	&	2.23	$\pm$	0.18	&	 110205A	&	1.98	&	51.72	 $\pm$	 0.03	&	3.6	$\pm$	0.2	&	1.80	 $\pm$	0.04	 \\
100413A	&	3.9	&	51.40	$\pm$	0.07	&	0.7	$\pm$	0.1	&	1.25	$\pm$	0.07	&	110213A	&	 1.46	&	51.07	$\pm$	0.03	 &	 1.6	$\pm$	0.6	&	1.83	$\pm$	 0.12	\\
100418A	&	0.6235	&	50.00	$\pm$	0.09	&	1.0	$\pm$	0.2	&	2.16	$\pm$	0.25	&	110422A	&	 1.77	&	52.43	$\pm$	 0.01	 &	 30.7	$\pm$	1.0	&	1.35	$\pm$	 0.0	\\
100424A	&	2.465	&	51.00	$\pm$	0.13	&	0.4	$\pm$	0.1	&	1.83	$\pm$	0.13	&	110503A	&	 1.613	&	50.99	$\pm$	 0.02	 &	 1.4	$\pm$	0.1	&	1.35	$\pm$	 0.06	\\
100425A	&	1.755	&	51.36	$\pm$	0.11	&	1.4	$\pm$	0.2	&	2.42	$\pm$	0.32	&	110715A	&	 0.82	&	51.98	$\pm$	 0.01	 &	 53.9	$\pm$	1.1	&	1.63	$\pm$	 0.03	\\
100513A	&	4.772	&	51.70	$\pm$	0.11	&	0.6	$\pm$	0.1	&	1.62	$\pm$	0.14	&	110731A	&	 2.83	&	52.31	$\pm$	 0.02	 &	 11.0	$\pm$	0.3	&	1.15	$\pm$	 0.05	\\
100615A	&	1.398	&	51.56	$\pm$	0.02	&	5.4	$\pm$	0.2	&	1.87	$\pm$	0.04	&	110801A	&	 1.858	&	51.16	$\pm$	 0.09	 &	 1.1	$\pm$	0.2	&	1.84	$\pm$	 0.10	\\
100621A	&	0.542	&	50.95	$\pm$	0.01	&	12.8	$\pm$	0.3	&	1.90	$\pm$	0.03	&	110808A	 &	1.348	&	50.48	$\pm$	 0.32	 &	0.4	$\pm$	0.2	&	2.32	$\pm$	 0.43	\\
100704A	&	3.6	&	52.36	$\pm$	0.04	&	4.3	$\pm$	0.2	&	1.73	$\pm$	0.06	&	110818A	&	 3.36	&	51.79	$\pm$	0.10	 &	 1.6	$\pm$	0.3	&	1.58	$\pm$	 0.11	\\
100728A	&	1.567	&	51.50	$\pm$	0.02	&	5.1	$\pm$	0.2	&	1.18	$\pm$	0.02	&	111008A	&	 5	&	52.91	$\pm$	0.07	 &	 6.4	$\pm$	0.7	&	1.86	$\pm$	 0.09	\\
100728B	&	2.106	&	51.70	$\pm$	0.08	&	3.5	$\pm$	0.5	&	1.55	$\pm$	0.14	&	111107A	&	 2.893	&	51.50	$\pm$	 0.09	 &	 1.2	$\pm$	0.2	&	1.49	$\pm$	 0.14	\\
100814A	&	1.44	&	51.18	$\pm$	0.04	&	2.5	$\pm$	0.2	&	1.47	$\pm$	0.04	&	111123A	&	 3.1516	&	51.53	$\pm$	 0.06	 &	 0.9	$\pm$	0.1	&	1.68	$\pm$	 0.07	\\
																															
111209A	&	0.677	&	49.76	$\pm$	0.10	&	0.5	$\pm$	0.1	&	1.48	$\pm$	0.03	&	120907A	&	 0.97	&	50.89	$\pm$	 0.07	 &	 2.9	$\pm$	0.4	&	1.73	$\pm$	 0.25	\\
111225A	&	0.297	&	49.09	$\pm$	0.06	&	0.7	$\pm$	0.1	&	1.70	$\pm$	0.15	&	120909A	&	 3.93	&	51.99	$\pm$	 0.15	 &	 1.39	$\pm$	0.06	&	1.8	$\pm$	 0.3	\\
111228A	&	0.714	&	51.24	$\pm$	0.02	&	12.4	$\pm$	0.5	&	2.27	$\pm$	0.06	&	120922A	 &	3.1	&	52.04	$\pm$	 0.05	 &	 2.0	$\pm$	0.2	&	2.09	$\pm$	 0.08	\\
111229A	&	1.3805	&	50.81	$\pm$	0.11	&	1.0	$\pm$	0.2	&	1.85	$\pm$	0.33	&	121024A	&	 2.298	&	51.30	$\pm$	 0.10	 &	 1.3	$\pm$	0.2	&	1.41	$\pm$	 0.22	\\
120118B	&	2.943	&	52.02	$\pm$	0.07	&	2.2	$\pm$	0.3	&	2.08	$\pm$	0.11	&	121027A	&	 1.773	&	51.17	$\pm$	 0.08	 &	 1.3	$\pm$	0.2	&	1.82	$\pm$	 0.09	\\
120119A	&	1.728	&	51.94	$\pm$	0.02	&	10.3	$\pm$	0.3	&	1.38	$\pm$	0.04	&	121128A	 &	2.2	&	52.23	$\pm$	 0.06	 &	 12.9	$\pm$	0.4	&	1.32	$\pm$	 0.18	\\
120326A	&	1.798	&	51.80	$\pm$	0.03	&	4.6	$\pm$	0.2	&	2.06	$\pm$	0.07	&	121201A	&	 3.385	&	51.65	$\pm$	 0.11	 &	 0.8	$\pm$	0.1	&	1.90	$\pm$	 0.21	\\
120327A	&	2.81	&	52.00	$\pm$	0.03	&	3.9	$\pm$	0.2	&	1.52	$\pm$	0.06	&	121211A	&	 1.023	&	50.57	$\pm$	 0.14	 &	 1.0	$\pm$	0.3	&	2.36	$\pm$	 0.26	\\
120404A	&	2.876	&	51.64	$\pm$	0.09	&	1.2	$\pm$	0.2	&	1.85	$\pm$	0.13	&	130131B	&	 2.539	&	51.18	$\pm$	 0.12	 &	 1.00	$\pm$	0.20	&	1.15	 $\pm$	0.20	 \\
120422A	&	0.28	&	49.01	$\pm$	0.18	&	0.6	$\pm$	0.2	&	1.19	$\pm$	0.24	&	130215A	&	 0.597	&	50.33	$\pm$	 0.13	 &	 2.5	$\pm$	0.7	&	1.59	$\pm$	 0.14	\\
120712A	&	4	&	52.01	$\pm$	0.06	&	2.4	$\pm$	0.2	&	1.36	$\pm$	0.08	&	130408A	&	 3.758	&	52.23	$\pm$	0.16	 &	 4.9	$\pm$	1.0	&	1.28	$\pm$	 0.26	\\
120714B	&	0.3984	&	49.14	$\pm$	0.13	&	0.4	$\pm$	0.1	&	1.52	$\pm$	0.17	&	130418A	&	 1.218	&	50.50	$\pm$	 0.18	 &	 0.6	$\pm$	0.2	&	2.07	$\pm$	 0.17	\\
120722A	&	0.9586	&	50.44	$\pm$	0.14	&	1.0	$\pm$	0.3	&	1.90	$\pm$	0.25	&	130420A	&	 1.297	&	51.34	$\pm$	 0.03	 &	 3.4	$\pm$	0.2	&	2.18	$\pm$	 0.05	\\
120724A	&	1.48	&	50.80	$\pm$	0.20	&	0.6	$\pm$	0.2	&	2.45	$\pm$	0.26	&	130427A	&	 0.34	&	51.93	$\pm$	 0.01	 &	 331.0	$\pm$	4.6	&	1.21	$\pm$	 0.02	\\
120729A	&	0.8	&	50.69	$\pm$	0.03	&	2.9	$\pm$	0.2	&	1.62	$\pm$	0.08	&	130427B	&	 2.78	&	51.92	$\pm$	0.08	 &	 3.0	$\pm$	0.4	&	1.64	$\pm$	 0.15	\\
120802A	&	3.796	&	52.31	$\pm$	0.06	&	3.0	$\pm$	0.2	&	1.84	$\pm$	0.10	&	130505A	&	 2.27	&	52.58	$\pm$	 0.05	 &	 30.0	$\pm$	3.1	&	1.18	$\pm$	 0.07	\\
120811C	&	2.671	&	52.17	$\pm$	0.03	&	4.1	$\pm$	0.2	&	2.04	$\pm$	0.06	&	130511A	&	 1.3033	&	50.78	$\pm$	 0.09	 &	 1.3	$\pm$	0.2	&	1.35	$\pm$	 0.31	\\
120815A	&	2.358	&	51.86	$\pm$	0.10	&	2.2	$\pm$	0.3	&	2.29	$\pm$	0.23	&	130514A	&	 3.6	&	52.20	$\pm$	 0.05	 &	 2.8	$\pm$	0.3	&	1.80	$\pm$	 0.05	\\
																															
130604A	&	1.06	&	50.40	$\pm$	0.13	&	0.8	$\pm$	0.2	&	1.51	$\pm$	0.12	&	140419A	&	 3.956	&	52.24	$\pm$	 0.03	 &	 4.9	$\pm$	0.2	&	1.21	$\pm$	 0.04	\\
130606A	&	5.91	&	52.46	$\pm$	0.09	&	2.6	$\pm$	0.2	&	1.52	$\pm$	0.12	&	140423A	&	 3.26	&	51.78	$\pm$	 0.05	 &	 2.1	$\pm$	0.2	&	1.33	$\pm$	 0.06	\\
130610A	&	2.092	&	51.29	$\pm$	0.06	&	1.7	$\pm$	0.2	&	1.27	$\pm$	0.08	&	140430A	&	 1.6	&	51.40	$\pm$	 0.06	 &	 2.5	$\pm$	0.2	&	2.00	$\pm$	 0.22	\\
130612A	&	2.006	&	51.49	$\pm$	0.11	&	1.7	$\pm$	0.3	&	2.06	$\pm$	0.25	&	140506A	&	 0.889	&	51.38	$\pm$	 0.04	 &	 10.9	$\pm$	0.9	&	1.68	$\pm$	 0.16	\\
130701A	&	1.155	&	51.82	$\pm$	0.02	&	17.1	$\pm$	0.7	&	1.58	$\pm$	0.05	&	140512A	 &	0.725	&	50.95	$\pm$	 0.02	 &	6.8	$\pm$	0.3	&	1.45	$\pm$	 0.04	\\
130831A	&	0.4791	&	50.85	$\pm$	0.02	&	13.6	$\pm$	0.6	&	1.93	$\pm$	0.05	&	140515A	 &	6.32	&	52.28	$\pm$	 0.10	 &	0.9	$\pm$	0.1	&	1.85	$\pm$	 0.13	\\
130907A	&	1.238	&	51.99	$\pm$	0.01	&	25.6	$\pm$	0.5	&	1.17	$\pm$	0.02	&	140518A	 &	4.707	&	52.11	$\pm$	 0.09	 &	1.0	$\pm$	0.1	&	1.97	$\pm$	 0.13	\\
130925A	&	0.347	&	50.24	$\pm$	0.04	&	7.3	$\pm$	0.6	&	2.09	$\pm$	0.04	&	140629A	&	 2.275	&	51.95	$\pm$	 0.06	 &	 4.2	$\pm$	0.4	&	1.86	$\pm$	 0.11	\\
131030A	&	1.293	&	52.10	$\pm$	0.01	&	28.1	$\pm$	0.7	&	1.30	$\pm$	0.03	&	140703A	 &	3.14	&	52.05	$\pm$	 0.11	 &	2.8	$\pm$	0.6	&	1.74	$\pm$	 0.13	\\
131103A	&	0.599	&	50.12	$\pm$	0.09	&	1.5	$\pm$	0.3	&	1.97	$\pm$	0.19	&	140710A	&	 0.558	&	50.15	$\pm$	 0.07	 &	 1.9	$\pm$	0.3	&	2.00	$\pm$	 0.23	\\
131105A	&	1.686	&	51.46	$\pm$	0.08	&	3.5	$\pm$	0.6	&	1.45	$\pm$	0.11	&	140907A	&	 1.21	&	51.05	$\pm$	 0.04	 &	 2.5	$\pm$	0.2	&	1.72	$\pm$	 0.07	\\
131117A	&	4.18	&	51.74	$\pm$	0.12	&	0.7	$\pm$	0.1	&	1.79	$\pm$	0.18	&	141004A	&	 0.57	&	50.68	$\pm$	 0.02	 &	 6.1	$\pm$	0.3	&	1.86	$\pm$	 0.08	\\
140206A	&	2.73	&	52.69	$\pm$	0.02	&	19.4	$\pm$	0.5	&	1.58	$\pm$	0.03	&	141026A	 &	3.35	&	51.54	$\pm$	 0.31	 &	0.4	$\pm$	0.2	&	2.34	$\pm$	 0.19	\\
140213A	&	1.2076	&	52.03	$\pm$	0.02	&	23.5	$\pm$	0.8	&	1.80	$\pm$	0.04	&	141109A	 &	2.993	&	51.86	$\pm$	 0.04	 &	2.5	$\pm$	0.2	&	1.52	$\pm$	 0.07	\\
140301A	&	1.416	&	50.70	$\pm$	0.16	&	0.7	$\pm$	0.2	&	1.96	$\pm$	0.28	&	141121A	&	 1.47	&	50.80	$\pm$	 0.18	 &	 0.9	$\pm$	0.3	&	1.73	$\pm$	 0.13	\\
140304A	&	5.283	&	52.04	$\pm$	0.07	&	1.7	$\pm$	0.2	&	1.29	$\pm$	0.08	&	141220A	&	 1.3195	&	51.61	$\pm$	 0.04	 &	 8.9	$\pm$	0.7	&	1.30	$\pm$	 0.08	\\
140311A	&	4.95	&	52.10	$\pm$	0.23	&	1.3	$\pm$	0.5	&	1.67	$\pm$	0.23	&	141221A	&	 1.452	&	51.33	$\pm$	 0.03	 &	 3.1	$\pm$	0.2	&	1.74	$\pm$	 0.08	\\
140318A	&	1.02	&	50.14	$\pm$	0.34	&	0.5	$\pm$	0.2	&	1.35	$\pm$	0.28	&	141225A	&	 0.915	&	50.45	$\pm$	 0.16	 &	 1.3	$\pm$	0.4	&	1.32	$\pm$	 0.15	\\
																															
150206A	&	2.087	&	52.08	$\pm$	0.02	&	10.1	$\pm$	0.4	&	1.33	$\pm$	0.04	&	150323A	 &	0.593	&	50.67	$\pm$	 0.02	 &	5.4	$\pm$	0.3	&	1.85	$\pm$	 0.07	\\
150301B	&	1.5169	&	51.30	$\pm$	0.03	&	3.0	$\pm$	0.2	&	1.47	$\pm$	0.07	&	150403A	&	 2.06	&	52.28	$\pm$	 0.02	 &	 17.6	$\pm$	0.6	&	1.23	$\pm$	 0.04	\\
150314A	&	1.758	&	52.45	$\pm$	0.01	&	38.5	$\pm$	0.9	&	1.08	$\pm$	0.03	&	150413A	 &	3.2	&	51.83	$\pm$	 0.10	 &	 1.6	$\pm$	0.3	&	1.75	$\pm$	 0.14	\\

\enddata

\tablenotetext{a}{The 1-sec peak isotropic luminosity.}
\tablenotetext{b}{The 1-sec peak photon flux .}
\tablenotetext{c}{All of the 1-sec peak spectra of bursts are fitted to a simple power law gives a photon index
 $\Gamma$ .}

\end{deluxetable}

\begin{deluxetable}{cccc}
\tablewidth{350pt} 
\tabletypesize{\footnotesize}
\tablecaption{Results of smoothly broken power-law function fits to the GRB cumulative luminosity distributions in given redshift ranges for GRBs in our sample.}
\tablehead{ \colhead{ Redshift Interval}&
\colhead{$\alpha_{1}$}& \colhead{$\alpha_{2}$}& \colhead{$L_{b}/10^{51}$ erg $s^{-1}$}}
\startdata
$0<z\leq1$         &  $1.02\pm0.04$ & $1.96\pm0.09$ & $0.30\pm0.10$                 \\
$1<z\leq2$         &  $1.19\pm0.03$ & $3.14\pm0.18$ & $4.49\pm0.52$                 \\
$2<z\leq3$         &  $1.19\pm0.10$ & $2.51\pm0.18$ & $6.35\pm2.29$                  \\
$3<z\leq9.4$       &  $0.98\pm0.10$ & $2.84\pm0.17$ & $7.51\pm1.46$                  \\

\enddata
\end{deluxetable}

\clearpage

\begin{figure*}
  \centering
\resizebox{12cm}{!}
{\includegraphics{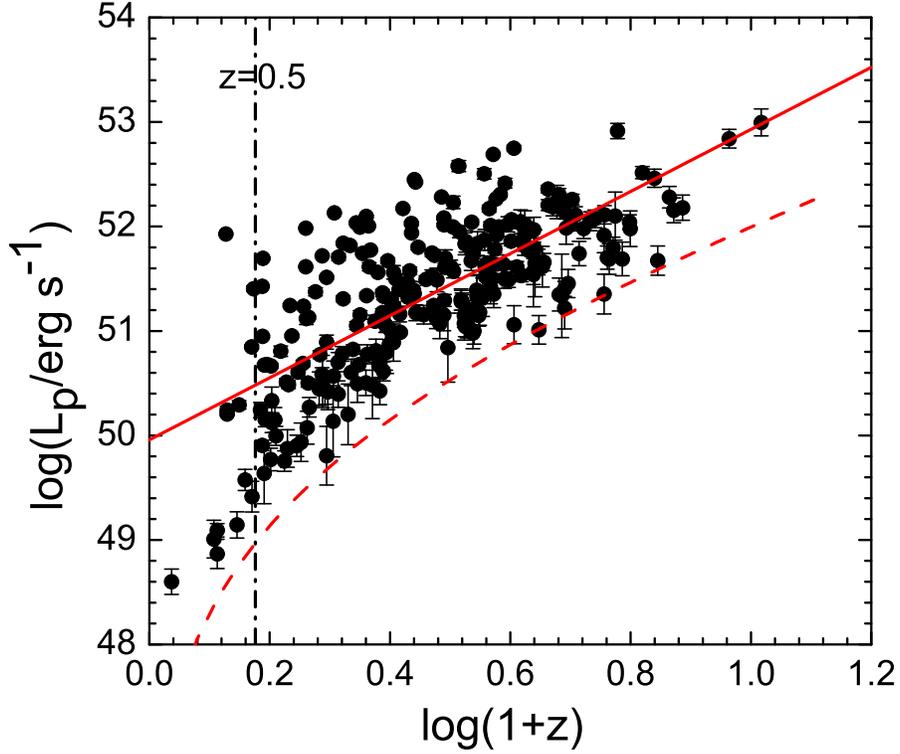}} \caption{Distribution of the GRBs in our sample in the $L_{\rm p}-z$ plane. The dashed line is the threshold for the BAT flux limit of $1.0\times 10^{-8}$ erg cm$^{-2}$ s$^{-1}$ (taken from http://swift.gsfc.nasa.gov/about\_swift/bat\_desc.html). The solid line is the best fit to data  with the ordinary least squares algorithm, $\log L_{\rm p}=(49.98\pm 0.09)+(2.95\pm 0.19)\log[(1+z)]$. The vertical dashed line marks $z=0.5$.}
\label{L-z}
\end{figure*}

\begin{figure}
  \centering
\resizebox{12cm}{!}{\includegraphics{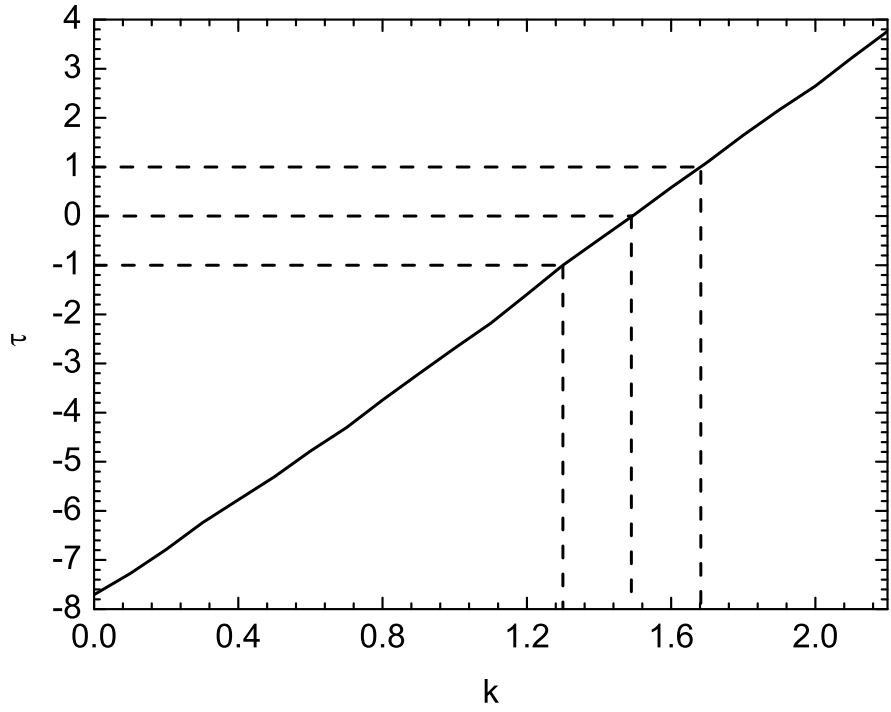} }
\caption{$\tau$ as a function of $k$ calculated with the $\tau$ statistical method for the GRBs in our sample assuming $L\propto{\rm }(1+z)^{k}$. The vertical and horizonal dashed lines mark the $k$ value that corresponds to $\tau=0$ in 1$\sigma$ confidence level, i.e., $k=1.49\pm0.19$.}  \label{tauz}
\end{figure}

\begin{figure}
  \centering
\resizebox{12cm}{!}{\includegraphics{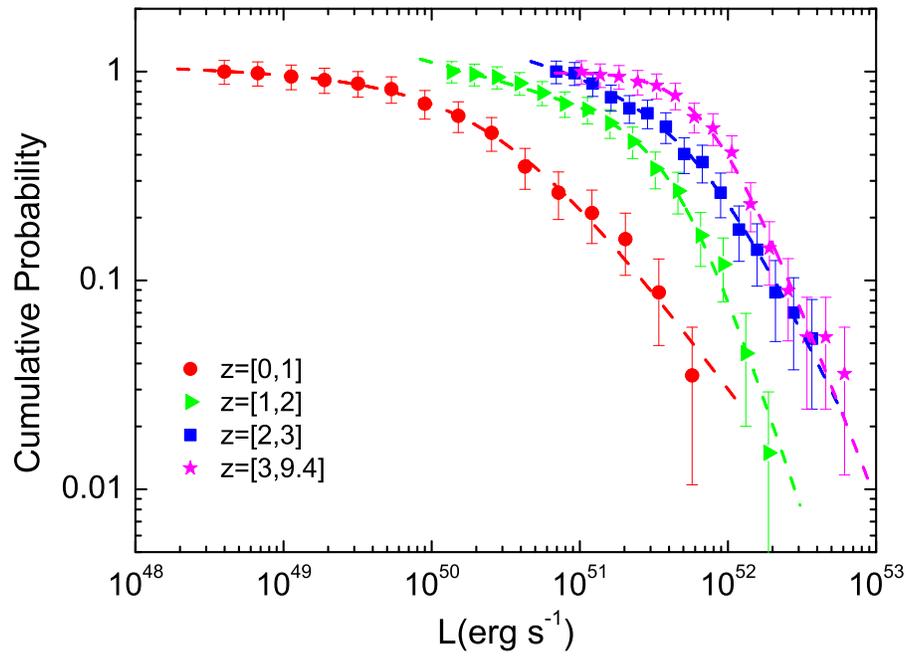}}
\caption{Cumulative luminosity distributions for different redshift intervals for the GRBs in our sample.
The lines represent the best fit to the data with a smooth broken power-law function.}
\label{CumuL}
\end{figure}

\begin{figure}
  \centering
\includegraphics[angle=0,scale=0.25]{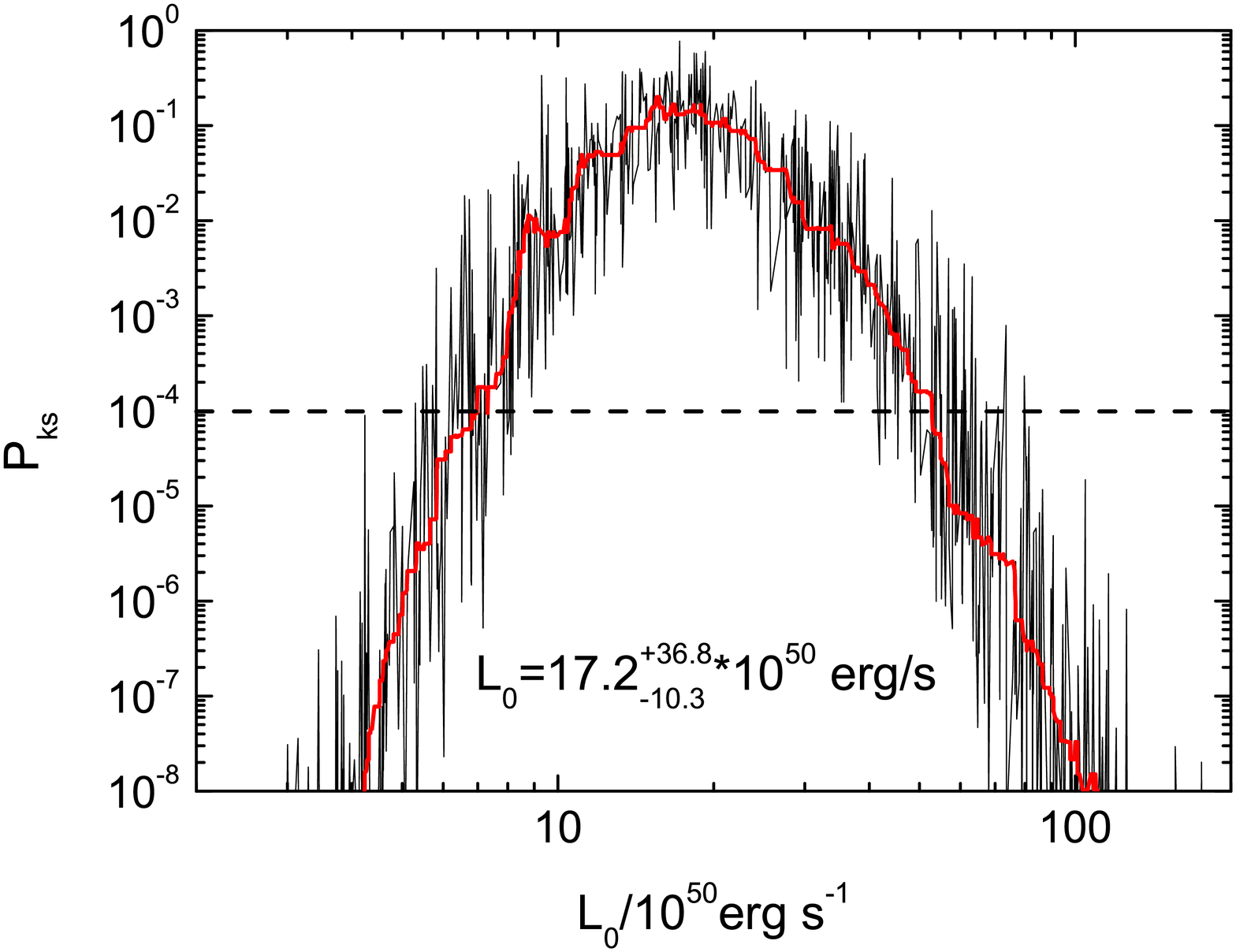}
\includegraphics[angle=0,scale=0.25]{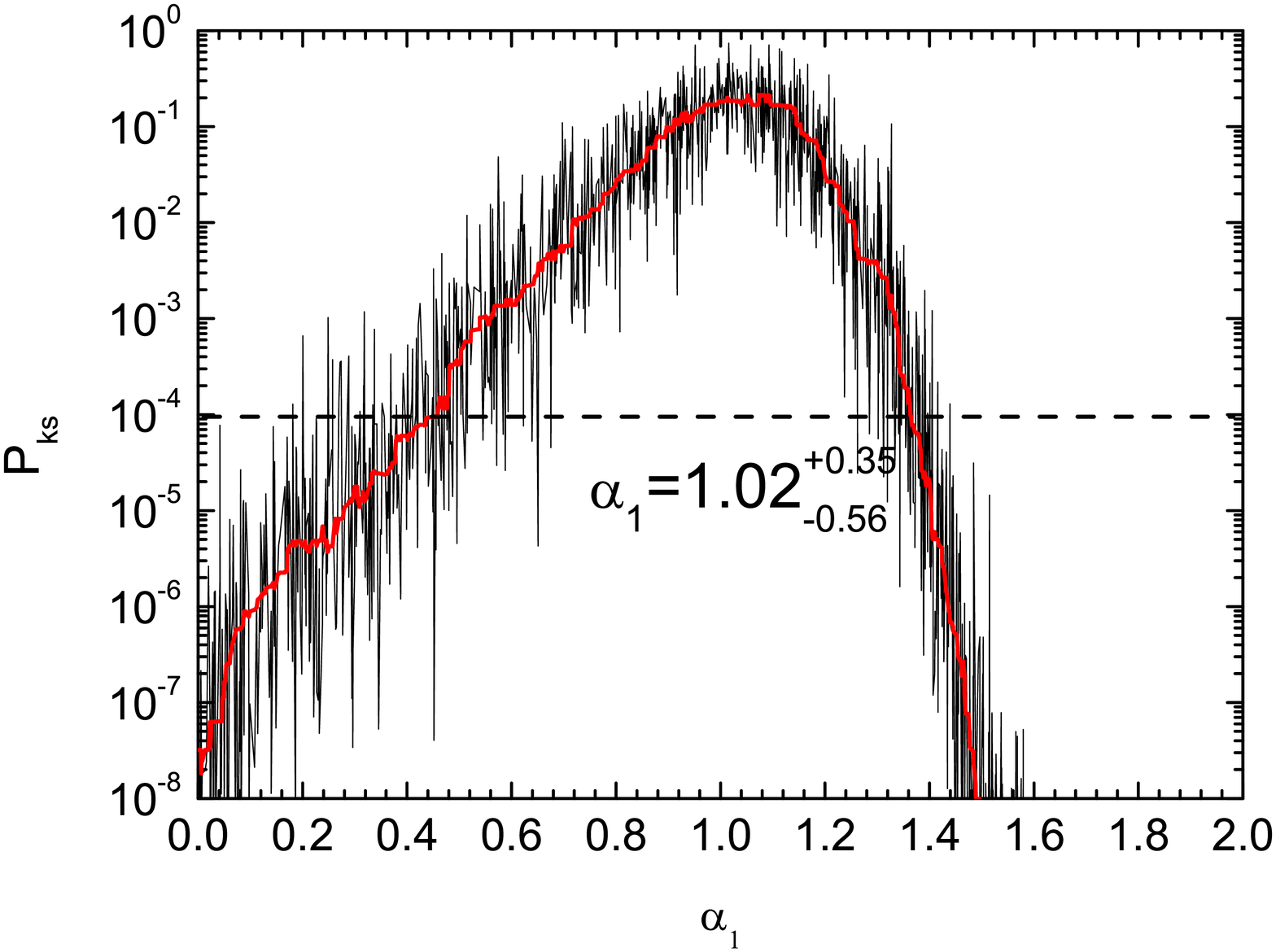}
\includegraphics[angle=0,scale=0.25]{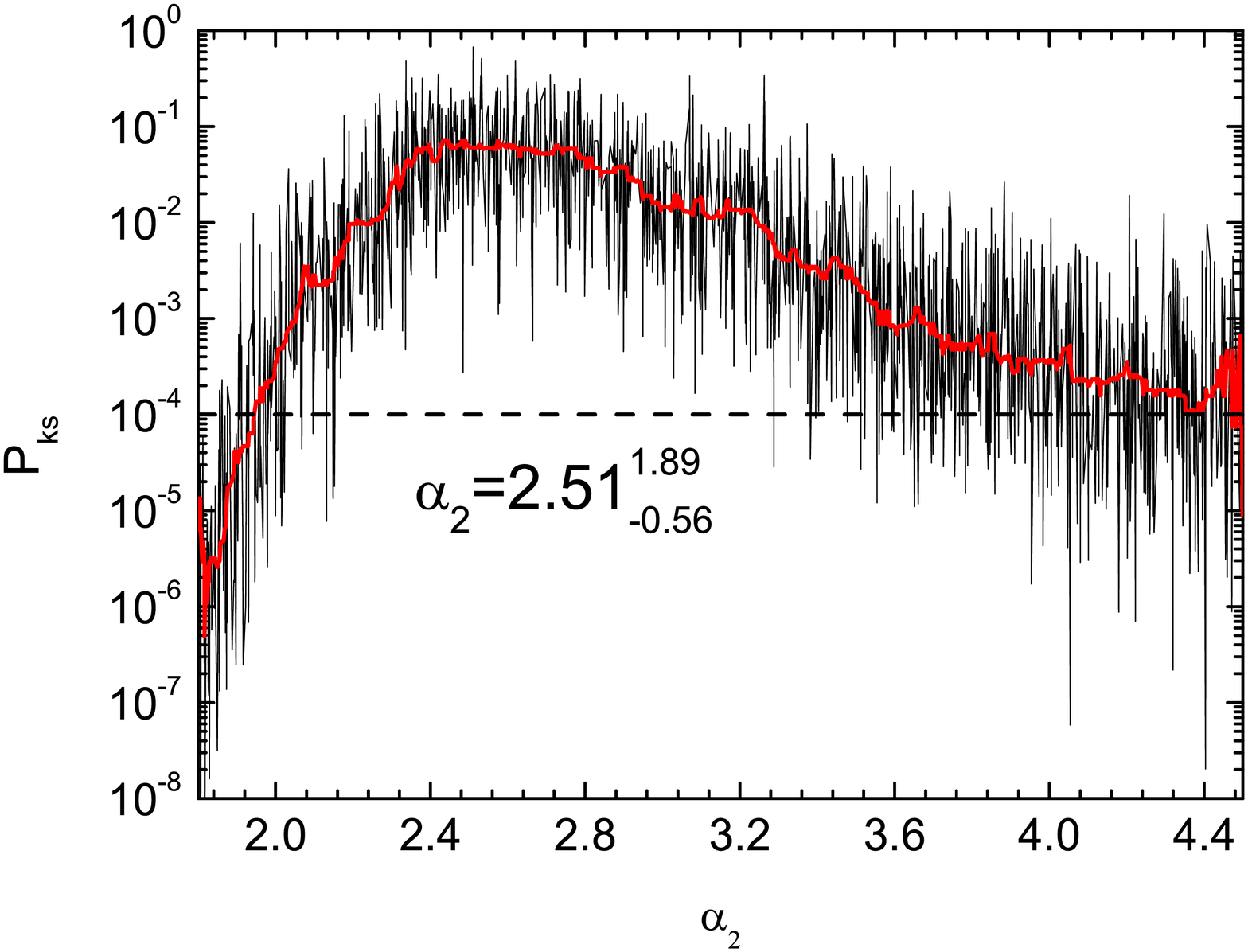}
\includegraphics[angle=0,scale=0.25]{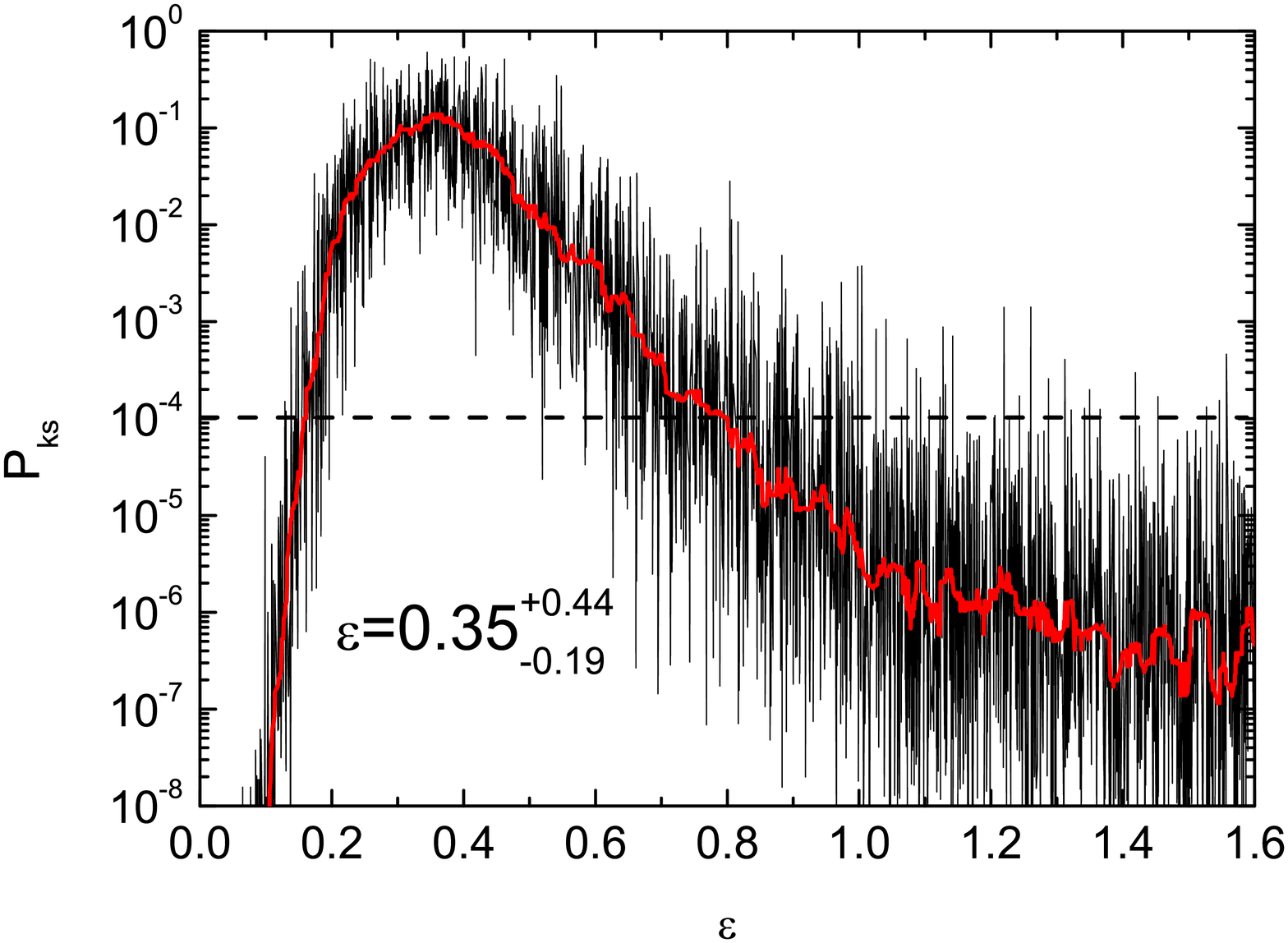}
\includegraphics[angle=0,scale=0.25]{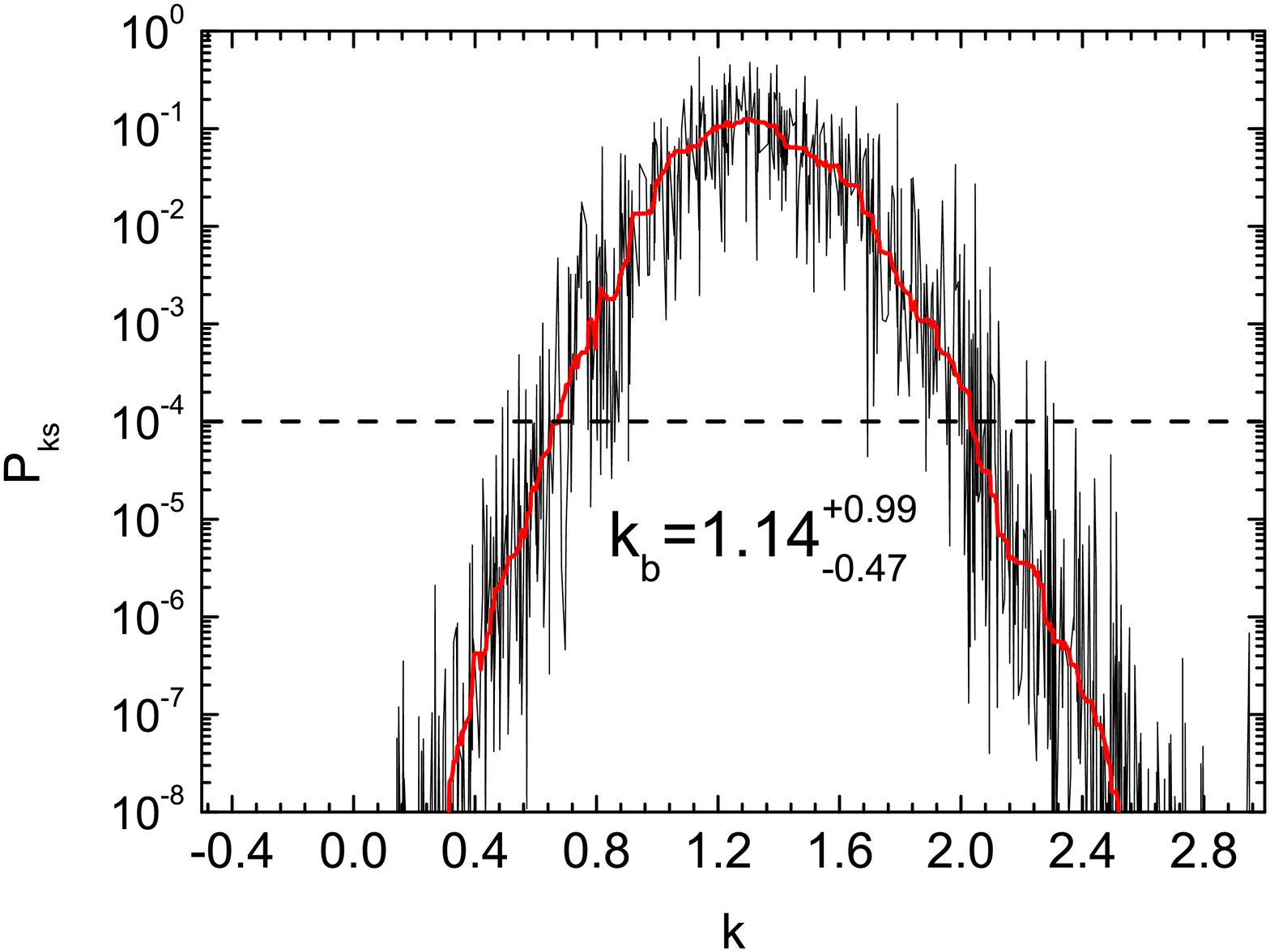}

\caption{Two-dimensional $P_{\rm KS}$ for the GRB distribution in the the $\log L_{\rm p}-\log (1+z)$ plane as a function of the model parameters along with the smoothed curves (red lines). The confidence levels of this parameters are measured with $P_{KS}>10^{-4}$, as marked with  horizonal lines.
 }  \label{MCdata}
\end{figure}

\begin{figure}
  \centering
\includegraphics[angle=0,scale=0.6]{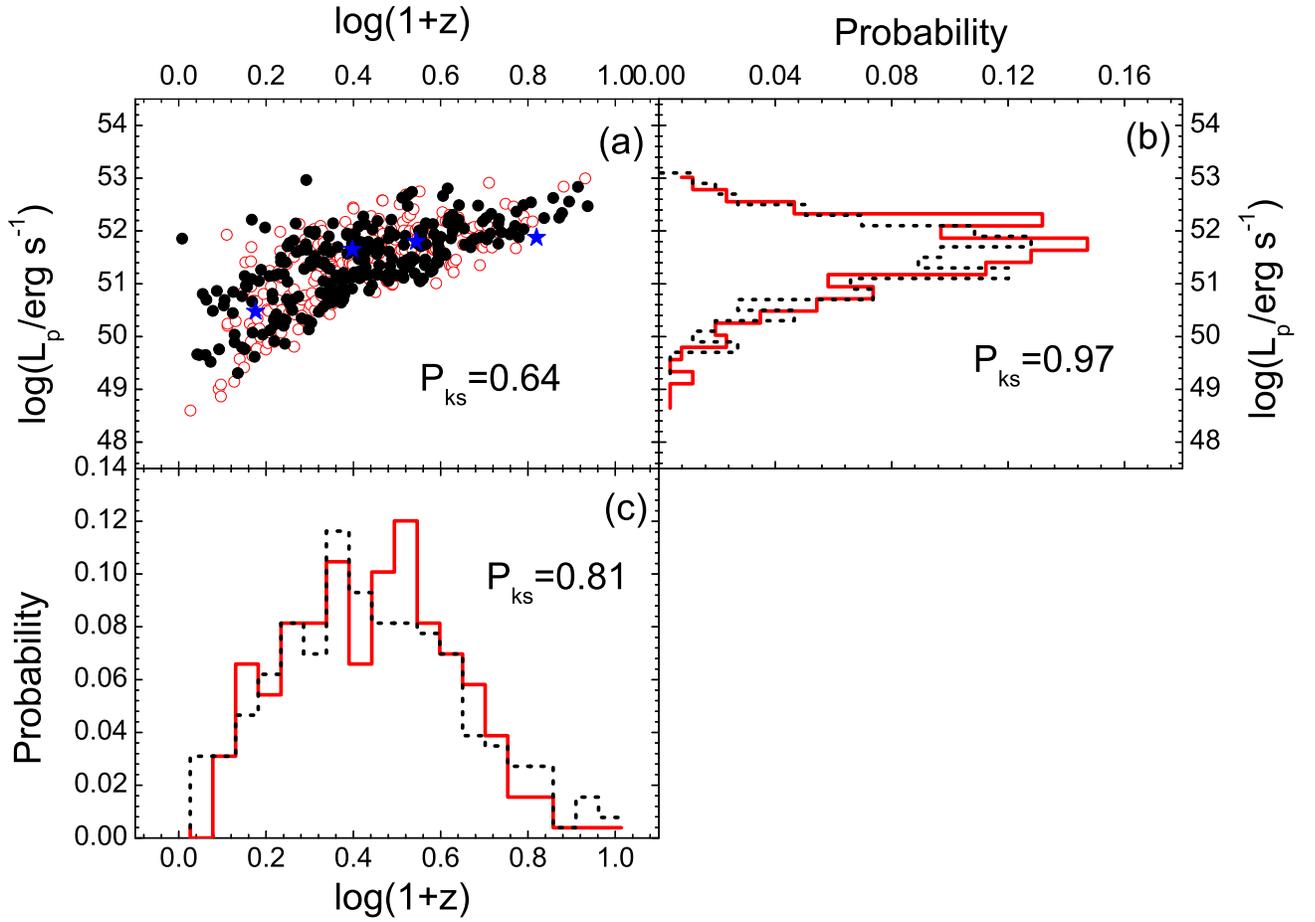}
\caption{Demonstration of the consistency between the observations (open dots and solid lines) and our simulations (solid  dots
and dashed lines) with the best parameter set derived in our simulations. The break luminosity $L_{\rm b}$ derived in different redshift ranges (blue stars) and the $p_{\rm KS}$ values are also marked.} \label{Lp_z}
\end{figure}

\begin{figure}
  \centering
\includegraphics[angle=0,scale=0.5]{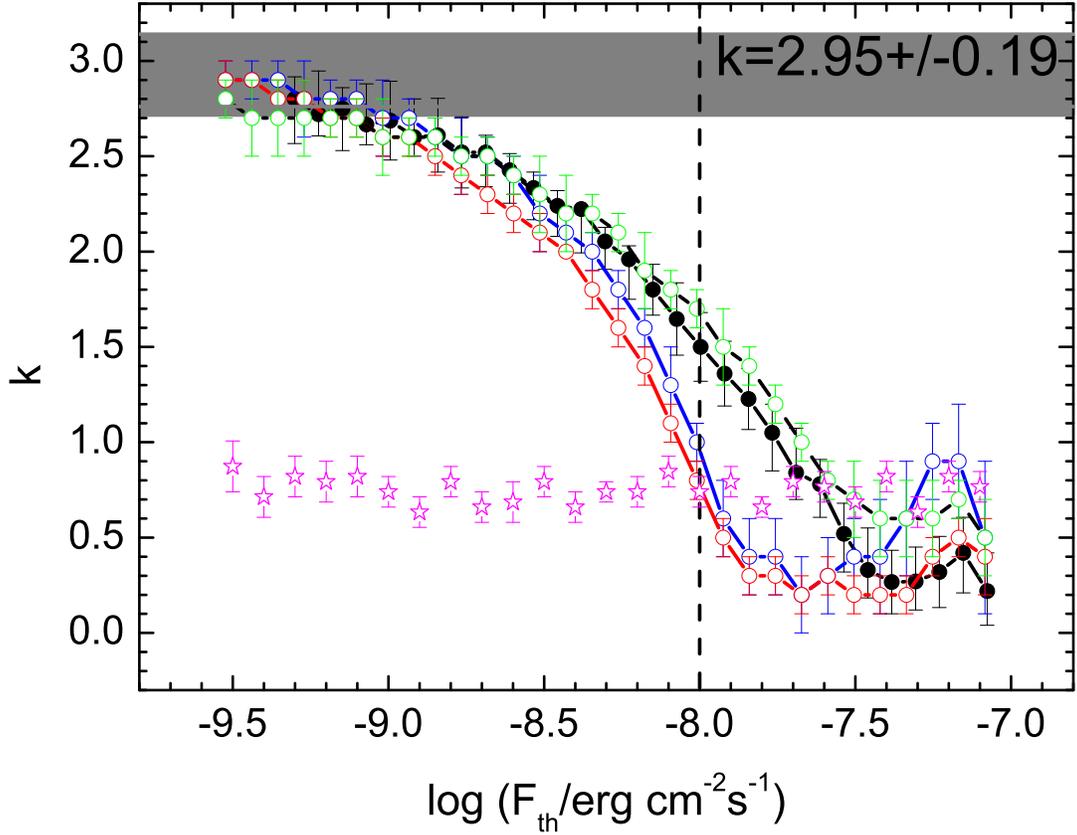}
\caption{$k$ value derived from the $\tau$ statistics method as a function of the adopted flux threshold for different samples: {\em solid} dots for the observed BAT sample of 258 GRBs; {\em open} green, blue, and red dots for a mock observation sample (258 GRBs), a mock {\em complete} BAT sample of 258 GRBs and a mock {\em complete} BAT sample of 1000 GRBs for the BAT threshold ($F_{\rm th}=1\times 10^{-8}$ erg cm$^{-2}$ s$^{-1}$), respectively; Open stars for the mock {\em complete} samples of 1000 GRBs in different thresholds and deriving $k$ values from these samples with the corresponding threshold. The grey band shows the $k$ value directly derived from the best fit to the data without considering any truncation effect, which should be the upper limit of the $k$ value. The vertical dashed line marks the BAT threshold. All mock GRB samples are generated from our simulations with the best parameter set as reported in the main text. For generating a mock {\em complete} sample, biases of GRB trigger and redshift measurement are removed in our simulations.}  \label{k_threshold}
\end{figure}

\label{lastpage}

\end{document}